\documentclass[reprint,aps,prc,superscriptaddress,floatfix]{revtex4-2}

\usepackage{graphicx}
\usepackage{amsmath,amssymb,amsfonts}
\usepackage{dcolumn}
\usepackage{bm}
\usepackage[dvipsnames]{xcolor}
\usepackage{lmodern}
\usepackage{booktabs}
\usepackage{multirow}
\usepackage[colorlinks=true,linkcolor=blue,citecolor=blue,urlcolor=blue]{hyperref}
\usepackage{orcidlink}

\begin{document}

\title{\boldmath Observation of a dominant $0f_{7/2}$ neutron configuration in the $^{32}$Si $J^{\pi}=5^-$ isomeric state}

\author{C.~R.~Hoffman\orcidlink{0000-0001-7141-9827}}
\email{Contact author: crhoffman@anl.gov}
\affiliation{Physics Division, Argonne National Laboratory, Lemont, IL 60439, USA}

\author{G.~L.~Wilson}
\affiliation{Department of Physics and Astronomy, Louisiana State University, Baton Rouge, LA 70803, USA}

\author{J.~Chen}
\altaffiliation{Present address: Department of Physics, Southern University of Science and Technology, Shenzhen, 518055, Guangdong, China}
\affiliation{Physics Division, Argonne National Laboratory, Lemont, IL 60439, USA}

\author{B.~P.~Kay}
\affiliation{Physics Division, Argonne National Laboratory, Lemont, IL 60439, USA}

\author{T.~L.~Tang}
\affiliation{Physics Division, Argonne National Laboratory, Lemont, IL 60439, USA}

\author{S.~R.~Carmichael\orcidlink{0000-0002-7454-1876}}
\altaffiliation{Present address: Physics Division, Argonne National Laboratory, Lemont, IL 60439, USA}
\affiliation{Department of Physics and Astronomy, University of Notre Dame, Notre Dame, IN 46556, USA}

\author{M.~Gott}
\altaffiliation{Present address: Oak Ridge National Laboratory, Oak Ridge, TN 37831, USA}
\affiliation{Physics Division, Argonne National Laboratory, Lemont, IL 60439, USA}

\author{S.~Lesher\orcidlink{0000-0002-5531-2867}}
\altaffiliation{Present address: Department of Physics, North Carolina A\&T State University, Greensboro, North Carolina 27411, USA}
\affiliation{Department of Physics and Astronomy, University of Wisconsin - La Crosse, La Crosse, WI 54601, USA}

\author{M.~S.~Martin\orcidlink{0000-0003-0707-9607}}
\affiliation{Physics Division, Argonne National Laboratory, Lemont, IL 60439, USA}

\author{G.~E.~Morgan}
\altaffiliation{Present address: Cyclotron Institute, Texas A\&M University, College Station, TX 77843, USA}
\affiliation{Department of Physics and Astronomy, Louisiana State University, Baton Rouge, LA 70803, USA}
\affiliation{Physics Division, Argonne National Laboratory, Lemont, IL 60439, USA}

\author{J.~Wu}
\altaffiliation{Present address: Brookhaven National Laboratory, Upton, New York 11973, USA}
\affiliation{Physics Division, Argonne National Laboratory, Lemont, IL 60439, USA}

\date{\today}

\begin{abstract}
An yrast, $J^{\pi}=5^-$, spin-trap isomer has been previously identified in $^{32}$Si. The isomeric state decays predominantly via a hindered $E3$ transition [B($E3$) = 0.0841(10)~W.u.], bypassing a nearby $E2$ decay path to the first excited $3^-$ level. The single-neutron aspects of these negative parity levels were investigated via the $^{31}$Si$(d$,$p)^{32}$Si reaction at 9.6~MeV/$u$ using HELIOS and the ATLAS in-flight facility. The $5^-$ state appears as a dominant $\ell=3$ transfer with a relatively large spectroscopic factor, confirming its single-particle $\nu0f_{7/2}$ character. The yrast $3^-$ level had a reduced $\ell=3$ spectroscopic factor of $\approx$ 0.44 compared to that of the $5^-_1$ level. This is similar to the situation observed in nearby $^{34}$S which by contrast has a measured B($E2, 5^-\rightarrow 3^-$) transition strength closer to 1~W.u.. It has been concluded that the hinderance of the $5^-_1\rightarrow 3^-_1$ transition in $^{32}$Si is not primarily due to the differing overlaps in the neutron structure. Instead, the lack of participation by both the protons and the neutrons in the transition is proposed as the transition-strength reduction mechanism.
\end{abstract}

\maketitle

\section{Introduction}\label{sec:intro}

The $Z=14$, $N=18$ nucleus $^{32}$Si was discovered in 1953 and its excited-state level structure has been experimentally studied through various complementary methods~\cite{ref:Lindner1953,ref:Thoennessen2012,ref:For82,ref:Fornal1997,ref:Pro72,ref:Gui74,ref:Paul2001,ref:Heery2024,ref:Williams2023,ref:Williams2025,ref:Williams2025b}. At intermediate excitation energies, the $^{32}$Si nucleus has an yrast spin-trap nanosecond isomeric state \textemdash a rare occurrence in even-even $1s-0d$-shell nuclei. The state has been definitively assigned a spin-parity of $J^\pi = 5^-$ and determined to exhibit a mean lifetime ($\tau_{\text{mean}}$) of $46.9(5)$ ns~\cite{ref:Williams2023,ref:Williams2025,ref:Williams2025b}. The isomer decays predominantly ($>99$\%) via a high-energy electric-octupole ($E3$) transition of $3562.84(14)$ keV to the $2^+_1$ first-excited state. The reduced transition probability of this $E3$ decay, $B(E3; 5^- \to 2^+_1) = 0.0841(10)$ W.u., is significantly hindered compared to typical $3^-_1\rightarrow 0^+_1$ and other $5^-_1\rightarrow 2^+_1$ $E3$ transitions in this mass region~\cite{ref:Kibedi2002,ref:Greene1972}.

The present manuscript reports on an experimental study of the excited levels in $^{32}$Si over an excitation range of $E_x\approx3.5 - 8.5$~MeV using the ($d$,$p$) reaction at 9.6~MeV/$u$. This approach provides the single-neutron character of states via overlap strengths or spectroscopic factors ($C^2S$) with the $^{31}$Si ground state. The $^{31}$Si ($N=17$) ground state has $J^{\pi}=3/2^+$ due to an unpaired neutron in the $0d_{3/2}$ orbital. The state is well described within the $1s-0d$ single-particle space, as evidenced by a relatively large $\ell=2$ normalized single-neutron overlap (spectroscopic factor) of $C^2S \gtrsim 0.8$ with the $^{30}$Si ground state (See Table 4 of Ref~\cite{ref:Piskor2000} and references within).  The addition of a neutron onto the 3/2$^+$ ground state affords access to states in $^{32}$Si with ranges in total angular momenta: $J = 0^+$ or $2^+$ from a neutron in the $1s_{1/2}$ or $0d_{3/2}$ orbitals, $J = 2 - 5^-$ from a neutron located in the $0f_{7/2}$ orbital, and $J = 0 - 3^-$ from a neutron located in the $0p_{3/2}$ orbital.

\section{Background}\label{sec:background}

\begin{figure}[htb]
    \centering
    \includegraphics[width=0.48\textwidth]{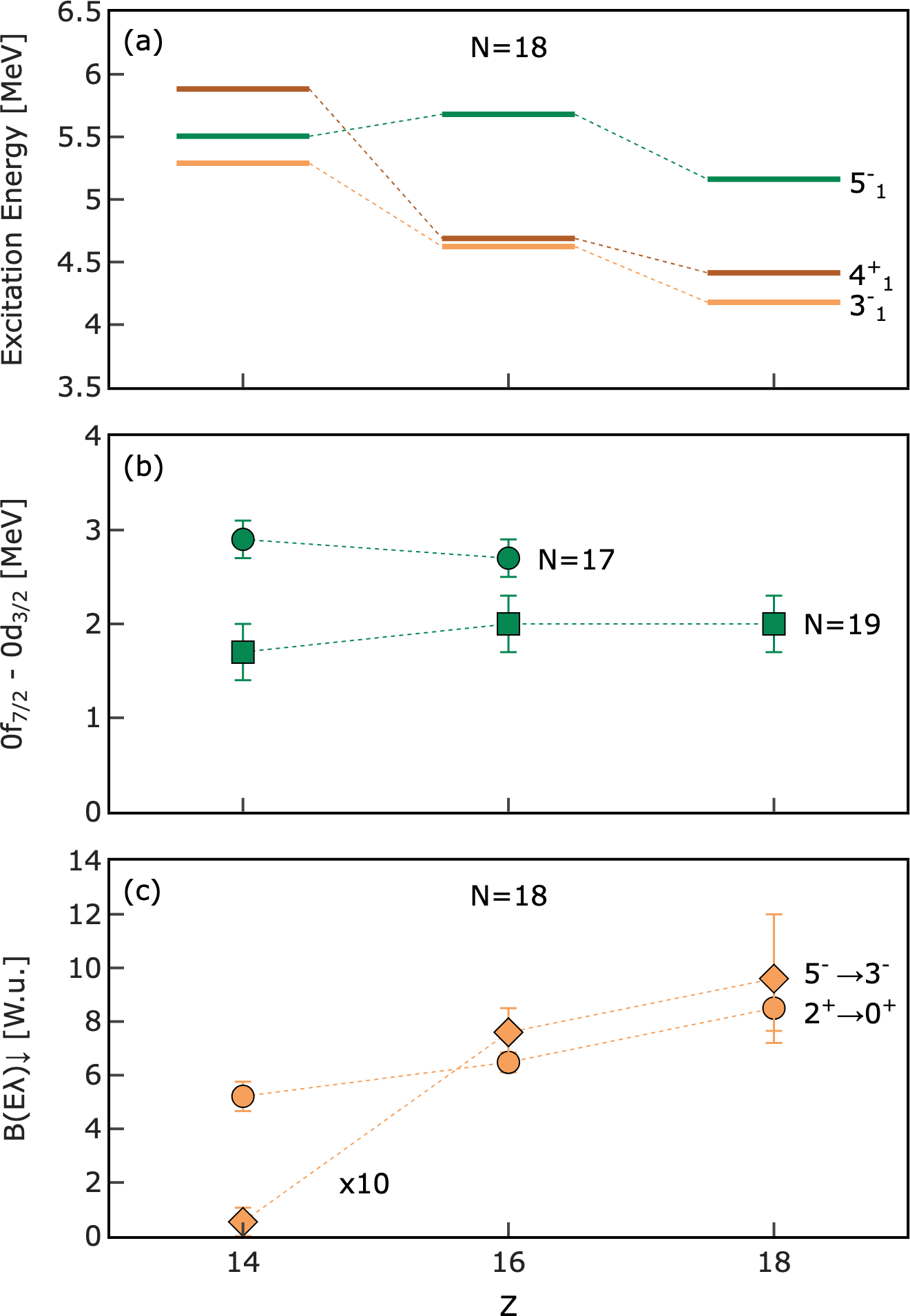}\\
    \caption{(a) The excited-state energies of the yrast $3^-$, $5^-$, and $4^+$ levels for the $N=18$ isotones of $^{32}$Si ($Z=14$), $^{34}$S ($Z=16$) and $^{36}$Ar ($Z=18$). (b) The $0f_{7/2} - 0d_{3/2}$ single-neutron energy centroid differences in the odd-$N$, $N=17$ and $N=19$ isotones~\cite{ref:Burgunder2014,ref:Eckle1989,ref:Chen2024,ref:Kuchera2024,ref:Mermaz1971}. (c) The $2^+_1$ to ground state [B($E2,2^+_1\rightarrow0^+_1$)] and $5^-_1$ to $3^-_1$ [B($E2,5^-_1\rightarrow3^-_1$)] quadrupole transition strengths in the $N=18$ even-even systems in Weisskopf units (W.u.)~\cite{ref:Raman2001,ref:Greene1972}. Note that the B($E2, 5^- \rightarrow 3^-$) include a factor of $\times10$.
    }
    \label{fig:sys}
\end{figure}

The appearance of a $5^-$ spin-trap isomeric state in $^{32}$Si at $E_x=5.505$~MeV is a consequence of the yrast $4^+_1$ level residing above the higher-$J$ level ($E_x=5.881$~MeV) as well as the close proximity (and configuration) of the yrast $3^-_1$ level ($E_x=5.288$~MeV). The properties leading to this energy pattern are conceptually understood from data on $^{32}$Si and the neighboring nuclei. To assist in the illustration of these points, the corresponding states in the $N=18$ isotones of $^{32}$Si, $^{34}$S, and $^{36}$Ar are plotted in Fig.~\ref{fig:sys}(a) for the yrast $3^-_1$, $5^-_1$ and $4^+_1$ excitation energies~\cite{ref:Williams2023}. Additional pertinent information are plotted in Figs.~\ref{fig:sys}(b) and (c), including the relative $0f_{7/2} - 0d_{3/2}$ single-neutron energy differences for select $N=17$ and $N=19$ isotones and dynamic quadrupole transition strengths, B($E2$)$\downarrow$, for both $2^+_1\rightarrow 0^+_1$ and $5^-_1\rightarrow 3^-_1$ transitions, respectively.

The $5^-$ excitation energies, $E_x$, remain relatively constant across these select $Z=14,16$ and 18, $N=18$ isotones. In the single-particle limit, the $5^-_1$ level is generated by a neutron cross-shell excitation configuration $(0d_{3/2})^{1}(0f_{7/2})^1$. The single-neutron adding data into $^{34}$S reinforces this picture via a large single-neutron overlap, $C^2S$, to its $5^-$ level~\cite{ref:Crozier1972,ref:VanDerBaan1971}. The implications are that the $E_x$ of the $5^-$ levels will then track the relative $0f_{7/2} - 0d_{3/2}$ single-neutron energy differences. This holds for the single-neutron centroids across some of the $N=17$~\cite{ref:MacGregor2021} and $N=19$~\cite{ref:Chen2024,ref:Kuchera2024} isotones as shown in Fig.~\ref{fig:sys}(b). The differences change by less than a few hundred keV over this region, consistent with the $5^-$ $E_x$ pattern. As discussed below, the results of the present work confirm the dominant $(0d_{3/2})^{1}(0f_{7/2})^1$ neutron configuration in the $5^-$ level of $^{32}$Si.

The energy of the $4^+_1$ state behaves inversely proportional to proton number ($Z$), ultimately rising above the $5^-_1$ $E_x$ in $^{32}$Si [Fig.~\ref{fig:sys}(a)]. A primary cause for this energy evolution is the required proton participation within the $1s-0d$ shell in order to generate a $J=4$ spin-state. In the case of $^{32}$Si, proton re-arrangement is hindered in comparison with the larger $Z$ isotopes due to its robust $Z=14$ sub-shell gap ($0d_{5/2}-1s_{1/2}$). Evidence for the sub-shell gap arises from a number of measurements including recent $^{32}$Si occupancy data and the 0.8(3) extracted proton vacancy of the $0d_{5/2}$ orbital~\cite{ref:Watwood2025}. Similarly, the sub-shell gap has been inferred from the reduced B($E2, 2^+_1\rightarrow 0^+_1$) transition strength in $^{32}$Si - reduced compared to the same transition strengths in $^{34}$S and $^{36}$Ar, as well as a possible reduction of the B($E2, 2^+_1\rightarrow 0^+_1$) transition strength relative to $1s-0d$-constrained shell-model predictions~\cite{ref:Heery2024,ref:Williams2023,ref:Williams2025,ref:Williams2025b}. Finally, as protons fill the $1s-0d$ proton orbitals for $Z>14$, excitations of protons into the $0f-1p$ shell becomes more accessible and contributes to the wave functions of $4^+$ state as noted in Ref.~\cite{ref:Williams2023}. No single-neutron information was obtained on the $4^+_1$ in the present work due to lack of direct access in the ($d$,$p$) reaction on the $3/2^+$ ground state.

The $E_x$ of the yrast $3^-_1$ levels increase with the removal of protons from the $1s-0d$ orbital [Fig.~\ref{fig:sys}(a)]. An independent particle estimation for the lowest $3^-$ level places it as part of the $(0d_{3/2})^{1}(0f_{7/2})^1$ spin-coupled multiplet. However, it is well established that the yrast $3^-$ levels in even-even systems are highly susceptible to couplings with coherent (octupole) correlations~\cite{ref:Crozier1972,ref:Bohr1975,ref:Spear1989,ref:Kibedi2002}. The larger admixtures in the wave functions of these levels result in reduced excitation energies and reduced single-neutron overlaps relative to other states in the multiplet. This is observed in $^{34}$S where the $^{33}$S($d$,$p$) overlaps show a $3^-_1$ to $5^-_1$ $C^2S$ ratio of $\approx0.25$~\cite{ref:VanDerBaan1971,ref:Crozier1972}. The rise in the $3^-_1$ $E_x$ with decreasing $Z$ is then interpreted as a reduction in the size or impact of such correlations [Fig.~\ref{fig:sys}]. The trends in reduced collectivity with decreasing $Z$ are in line with the trends in the quadrupole transition strengths B($E2, 2^+_1\rightarrow 0^+_1$), as well as the measured B($E3, 3^-_1 \rightarrow 0^+_1$) strengths in $^{34}$S [17(5) W.u.]and $^{36}$Ar [20(3) W.u.]~\cite{ref:Kibedi2002}, though large error bars persist for the latter.

Beyond the isomeric nature of the level, the $5^- \rightarrow 2^+_1$ decay transition out from the isomeric state showed a highly-hindered octupole transition strength of B($E3$, $5^-\rightarrow 2^+_1$) = 0.0841(10)~W.u.~\cite{ref:Williams2025}. Moreover, there is a notable by-passing of an accessible $E2$ decay path to the nearby $3^-_1$ level via a 0.217~MeV $\gamma$-ray transition~\cite{ref:Williams2025}. Based on a limit on the measured decay branch ratio, the B($E2, 5^-_1 \rightarrow 3^-_1$) transition strength in $^{32}$Si was placed at $<0.053$~W.u.~\cite{ref:Williams2025}. In $^{34}$S and $^{36}$Ar, the corresponding B($E2, 5^-_1 \rightarrow 3^-_1$) transition strengths are hindered but only on the order of $\approx0.5-1$~W.u.~\cite{ref:Greene1972}. As shown in Fig.~\ref{fig:sys}(c), typical first excited-state B($E2, 2^+_1\rightarrow 0^+_1$) transition strengths are an order of magnitude larger.

The cause of the highly-hindered $5^- \rightarrow 3^-$ transition strength in $^{32}$Si was discussed in Refs.~\cite{ref:Williams2023,ref:Williams2025b} as being indicative of the differences in the neutron configurations between the pure single-neutron $5^-$ level and the mixed (proton and neutron) $3^-_1$ level. The shell-model calculations in Refs.~\cite{ref:Williams2023,ref:Williams2025b}, in particular those based on the FSU interaction - an effective interaction developed specifically and successfully to describe particle-hole excitations from within the $1s-0d$ shells into the neighboring shell~\cite{ref:Lubna2020,ref:Lubna2024}, did reproduce the hindered B($E2, 5^-\rightarrow 3^-$) transition strength and calculated a change (reduction) in the $0f_{7/2}$ neutron occupancy of about 0.2 nucleons when going from the $5^-_1$ state to the $3^-_1$ state.

Tension arises with these arguments when considering the relative data and structures of $^{32}$Si and $^{34}$S. The $3^-_1$ level in $^{32}$Si should be the \emph{least} mixed of these $N=18$ isotones, but the B($E2, 5^-\rightarrow 3^-$) strength is by far the smallest. The $^{34}$S yrast $3^-$ level will have a larger difference in its $\nu0f_{7/2}$ occupancy relative to the $5^-_1$ level because of its increased mixing. A mismatch in the neutron wave functions may contribute to, but is not likely the primary cause of the highly-reduced B($E2, 5^-\rightarrow 3^-$) in $^{32}$Si. The $C^2S$ of the yrast $3^-$ level, relative to that of the $5^-_1$ level, was determined in the present work for $^{32}$Si to be $\approx0.44$.

\section{Measurement}\label{sec:meas}

The measurement was carried out at the ATLAS User Facility located at Argonne National Laboratory, USA. The short-lived $^{31}$Si beam was produced using the ATLAS in-flight system~\cite{ref:Hoffman2022}. The primary beam was $^{30}$Si at an energy 11.5~MeV/$u$ and a ($d$,$p$) production reaction took place on a cryo-cooled gas cell (90~K) filled to 1050~Torr with gaseous deuterium~\cite{ref:Rehm2011}. The resulting $^{31}$Si$^{14+}$ beam had an energy of 9.6~MeV/$u$, with a typical beam-on-target rate of $2 \times 10^4$~pps and an average $^{32}$Si purity of 30\%. The majority of the beam contamination was due to the unreacted $^{30}$Si$^{13+}$ primary beam which was accepted at an energy of $\sim$9~MeV/$u$. 

The HELIOS spectrometer~\cite{ref:Wuo07,ref:Lig10} measured outgoing protons in coincidence with $^{32}$Si recoils from the $^{31}$Si($d$,$p$) reaction at 9.6~MeV/$u$. The experimental setup, all contained within the 2.5~T solenoidal field, paralleled that shown in Fig. 2 of Ref.~\cite{ref:Hoffman2012}. A 150~$\mu$g/cm$^2$ CD$_2$ reaction target was located near the center of the magnetic field on the beam axis. The position-sensitive silicon detector (PSD) array was located upstream of the target covering a distance between -52~cm to -17~cm relative to the CD$_2$ target position and consisted of 30 individual detectors arranged evenly over six sides. The Si detectors measured total energies and positions of the protons. Because of the limited length of the PSD array, the strength of the magnetic field, and the reaction kinematics, the lowest-lying states were purposely excluded and only states above $E_x\approx3.5$~MeV in $^{32}$Si were accepted with sufficient angular coverage. The center-of-mass angular range for the $E_x = 5.505$~MeV, $J^{\pi}=5^-$ state was $20^{\circ}\lesssim \Theta_{cm}\lesssim40^{\circ}$.

The recoil detection took place $1100$~cm downstream of the CD$_2$ target location using four sets of $\Delta E-E$ Si detectors. Each set covered 1/4 of the azimuthal ($\phi$) angular range and had nominal thicknesses of $50~\mu$m for the $\Delta E$ and $1000~\mu$m for the $E$. The radial coverage was from an inner radius of 9~mm, so as to allow unreacted beam to pass through, to an outer radius of 48~mm. 

A background-free excitation energy spectrum (Fig.~\ref{fig:ex}) was constructed from the measured proton properties and the reaction kinematics for $^{32}$Si bound states. Protons were selected by the relative time difference between a proton signal in the silicon detector array and a recoil signal in one of the $\Delta E$ detectors. This relative time is proportional to the cyclotron period of a charged-particle orbit within the uniform solenoid field~\cite{ref:Wuo07}. The $^{32}$Si recoils of interest were selected by the combination of the $\Delta E$ and $E$ Si-detector information.

Angular distributions were constructed from detector yields subtending the same position along the PSD array. Each bin had the same solid angle coverage for a chosen $E_x$ and was a singular center-of-mass angle, $\Theta_{cm}$. The $\Theta_{cm}$ of each bin was taken as the average angle over the detector length as calculated from the reaction kinematics. The uncertainty on $\Theta_{cm}$ was $<0.5^{\circ}$ with the largest contributions coming from the detector-target positioning and the beam energy. An absolute scale on the differential cross sections was not determined.

A Distorted Wave Born approximation (DWBA) analysis was applied to the data using the finite-range software \texttt{PTOLEMY}~\cite{ref:Mac78}. The global optical model parameter sets of Refs.~\cite{ref:An06, ref:Koning2003} were applied for the deuteron and proton parameters, respectively. The sets of Refs.~\cite{ref:Perey1963d,ref:Perey1963p} were also explored, producing consistent results within the uncertainties. 
The deuteron bound-state wave function was defined through the Argonne $v_{18}$ potential~\cite{ref:Wiringa1995}. 
The neutron bound-state wave function for final states in $^{32}$Si utilized a Woods-Saxon potential. Parameters of $r_0 = 1.25$~fm, $a = 0.65$~fm, and a variable depth adjusted to match the final-state binding energy, were adopted for the central potential. Parameters of $V_{so} = 5.0$~MeV, $r_{so} = 1.1$~fm, and $a_{so} = 0.65$~fm were adopted for the spin-orbit potential. A systematic uncertainty of 10\% was applied to the $C^2S$ (relative) values to account for sensitives to the DWBA optical model and other parameters.

A check of the event selection procedures and DWBA analysis was done using ($d$,$p$) reaction data collected on the $^{30}$Si beam contaminant in comparison with previous work~\cite{ref:Piskor2000}. A $Q$-value resolution of $\sim$300~keV was achieved for the $^{30}$Si($d$,$p$) reaction. Known $\ell$ transfer values and $C^2S$ values were reproduced for states in $^{31}$Si below $E_x\approx4$~MeV. Uncertainties on the $C^2S$ were dominated by the statistics and resulting DWBA fits to the relative cross sections. 

\begin{table*}[ht!]
\centering
\caption{Spectroscopic factors $C^2S$ from the $^{31}$Si($d$,$p$) reaction at 9.6~MeV/$u$, listed by excitation energy in $^{32}$Si. All $C^2S$ are normalized to the 5505-keV state. The uncertainties on the $C^2S$ values include those from statistics, the DWBA fitting procedure, and those associated with the DWBA parameters.}
\label{tab:si32c2s}
\renewcommand{\arraystretch}{1.15}
\begin{tabular}{ccccccc}
\toprule
\multicolumn{2}{c}{$E_x$ [MeV]} & $J^{\pi}_f$ &\multicolumn{4}{c}{$C^2S$ (Expt.)} \\
\cmidrule(lr){1-2}\cmidrule(lr){3-3}\cmidrule(lr){4-7}
~\cite{ref:For82,ref:Fornal1997,ref:Pro72,ref:Gui74,ref:Paul2001,ref:Heery2024,ref:Williams2023,ref:Williams2025,ref:Williams2025b}
 &  This work & ~\cite{ref:For82,ref:Fornal1997,ref:Pro72,ref:Gui74,ref:Paul2001,ref:Heery2024,ref:Williams2023,ref:Williams2025,ref:Williams2025b}  & $\ell=0$ & $\ell=1$ & $\ell=2$ & $\ell=3$ \\
\midrule
4.231  & 4.26(3) & $2^+$      &  --        & --        & 0.48(20)\footnotemark[1] & --   \\
5.288  & 5.28(3) & $3^-$      &  --        & $<0.1$    & --       & 0.44(11)  \\
5.505  & 5.50(3) & $5^-$      &  --        & --        & --       & $\equiv1.00(11)$\footnotemark[2] \\
5.772  & 5.78(3)\footnotemark[3] & $3^{(-)}$  &  --        & 0.53$^{+28}_{-15}$  & --       & 0.38$^{+15}_{-28}$  \\
6.347  & 6.36(4) &$4^{(-)}$  &  --        & --          & --       & $<0.4$        \\
--  & 6.63(4) &-- & --       & --        & --       & --        \\
6.837  & \multirow{3}{*}{6.82(4)} & $(4^-,5^-)$ & --       & --        & --       & --        \\
6.850  &        & $(2,3)$     &  --        & --       & --        & -- \\
6.86(1)  &        & $(3^-)$     &  --        & --       & --        & -- \\
\bottomrule
\end{tabular}

\vspace{0.4em}
\begin{flushleft}
\footnotemark[1] Assumption of a pure $\ell=2$ neutron transfer, $\ell=0$ also allowed.\\
\footnotemark[2] Maximum $\ell=1$ contribution set at 10\%. \\
\footnotemark[3] Possibility that there are additional states present in this $E_x$ region.\\
\end{flushleft}
\end{table*}

\section{Results}\label{sec:res}

\begin{figure}[htb]
    \centering
    \includegraphics[width=0.48\textwidth]{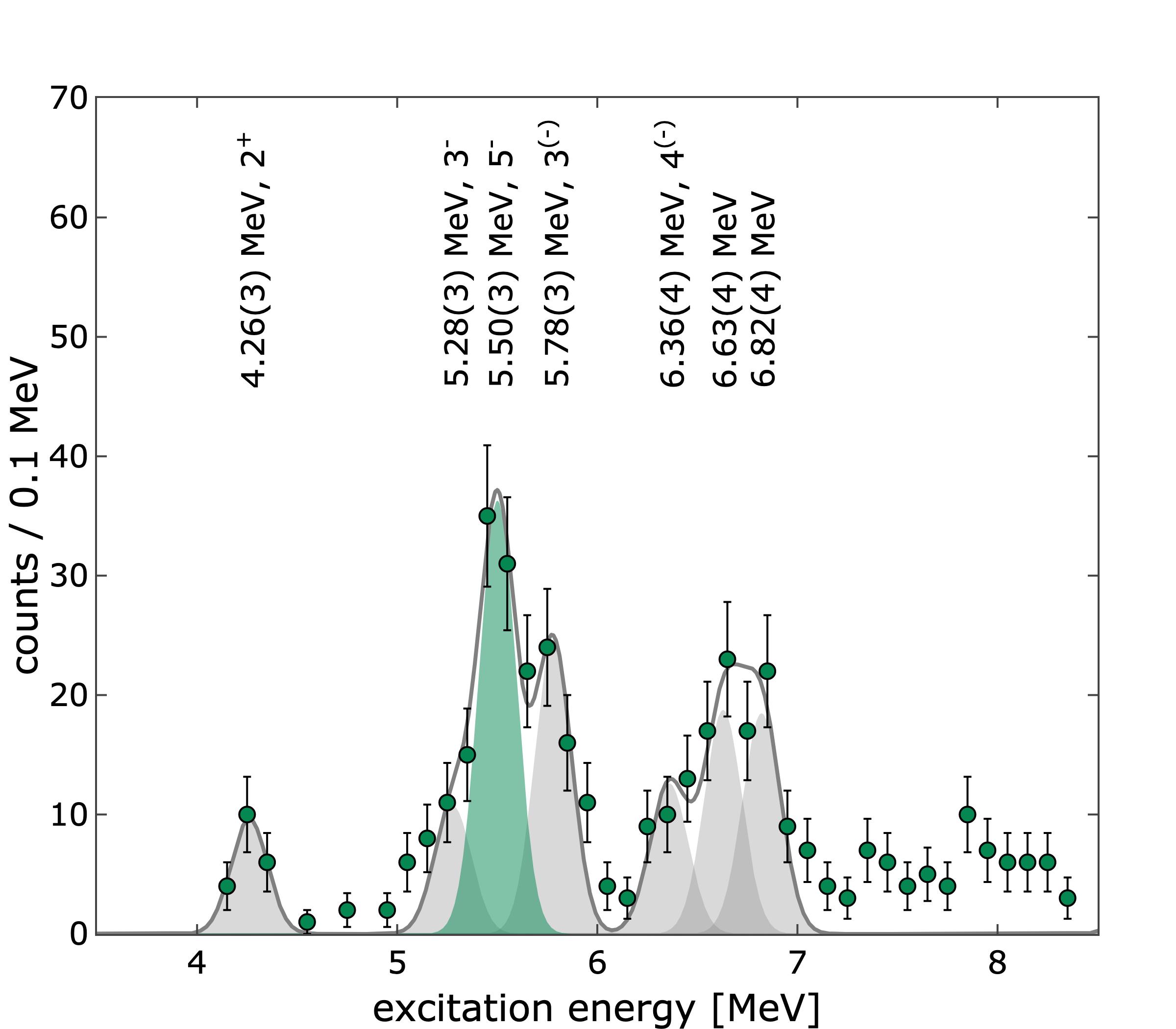}\\
    \caption{The $^{32}$Si excitation energy resulting from protons measured in the $^{31}$Si($d$,$p$)$^{32}$Si reaction. No corrections for the geometrical acceptances of the protons or recoils have been applied to the spectra. The centroid energies deduced from the displayed Gaussian fits are shown in MeV and their (suggested) association with previously identified levels are shown in Table~\ref{tab:si32c2s}. }
    \label{fig:ex}
\end{figure}

The excitation spectrum for $^{32}$Si generated from the $(d$,$p)$ reaction data is shown in Fig.~\ref{fig:ex} over the range of $E_x = 3.5 - 8.5$~MeV. The $E_x=5.505$~MeV $5^-$ state of interest is observed in the spectrum as the most abundant peak. A single Gaussian line-shape fit to this state using a fixed FWHM of 300~keV returned the appropriate centroid energy [$E_x = 5.50(3)$~MeV]~\cite{ref:Williams2025}. The fixed value for the FWHM was determined by the resolution obtained in the $^{30}$Si($d$,$p$) reaction data and confirmed by a single Gaussian fit to the known isolated 4.231-MeV state in the $^{32}$Si $E_x$ spectrum (Fig.~\ref{fig:ex}). 

Including Gaussian line-shapes for the previously identified states at 5.288~MeV ($3^-_1$) and 5.772~MeV [$3^{(-)}_2$] into the fit of the spectra gave a bulk reproduction of the yield between $E_x = 5 - 6$~MeV. Any definitive determination of other states populated in the reaction, for instance around the $E_x = 5.1-5.2$~MeV region, was outside the sensitivity of the available data. Also, the yrast $0^+$ and $2^+$ levels were not within the acceptance of this measurement and the $4^+$ level at 5.881~MeV was not accessible because of spin restrictions in the single-neutron transfer. There is a notable absence in the data of any sizable contributions to the yield around $E_x \approx  5.97$ MeV where an additional $3^-$ level was suggested via the ($t$,$p$) reaction~\cite{ref:For82}.

Over the $E_x = 6 - 7$~MeV region, only the results from the minimum number of Gaussian line-shapes required to reproduce the data are shown. The fit-results did coincide with the previously known $4^{(-)}$ state at $E_x = 6.347$~MeV as well as a number of other possible states previously observed around 6.8~MeV~\cite{ref:For82,ref:Williams2025}. It is likely that a number of states contribute to the yield observed around $E_x \gtrsim 7$~MeV. A complete summary of the energy-centroid fit results are listed in Table~\ref{tab:si32c2s} along with the known excitation energies.

\begin{figure}[!ht]
    \centering
    \includegraphics[width=0.48\textwidth]{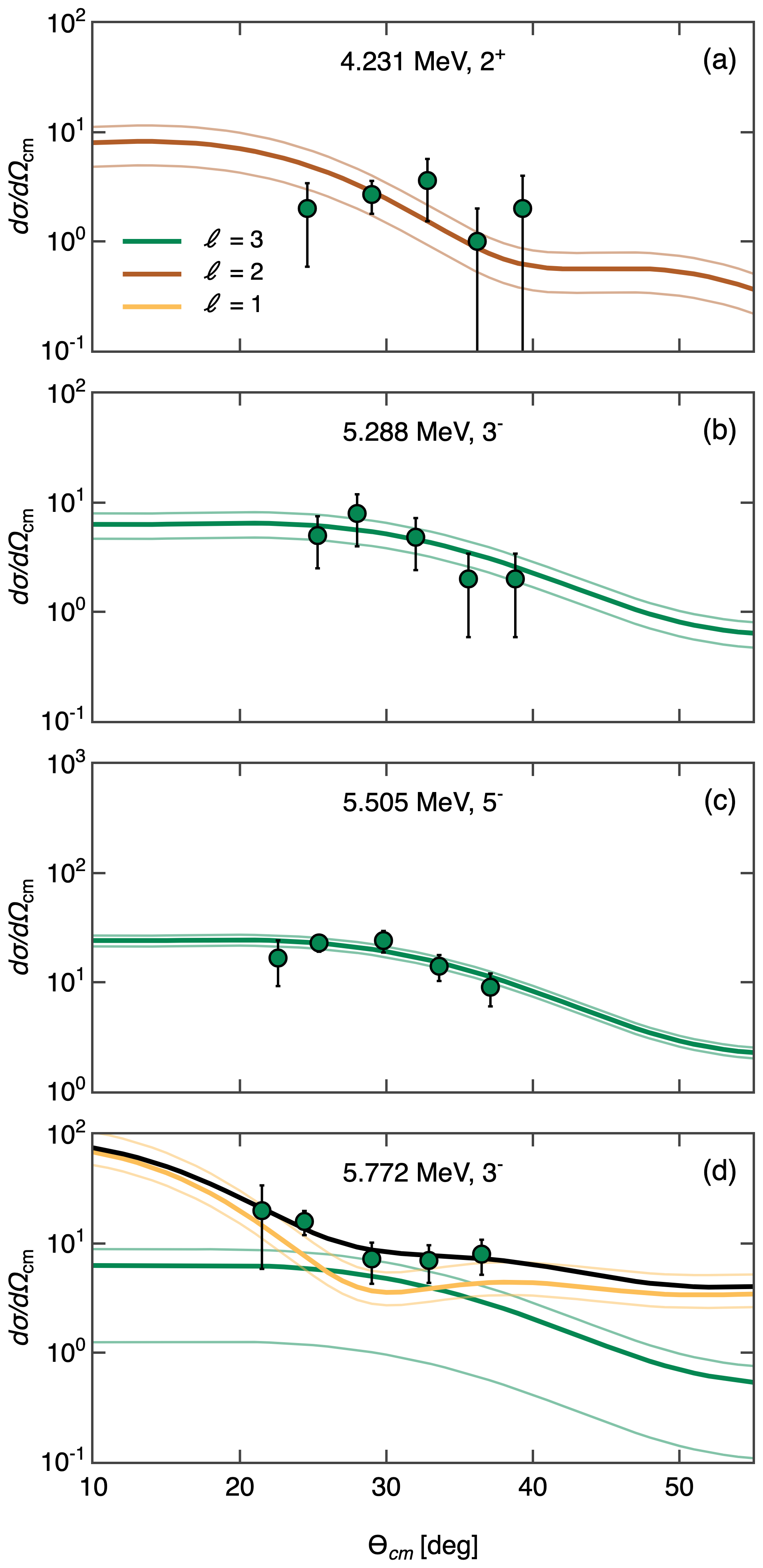}\\
    \caption{The data points show the measured angular distributions for select states in $^{32}$Si following the $^{31}$Si($d$,$p$)$^{32}$Si reaction. Solid lines result from the DWBA calculations described in the text and the properties listed in Table~\ref{tab:si32c2s}. The thick lines represent the best fits to the data and determine the $C^2S$, while the thin lines show the 1-$\sigma$ error limits of the fits. }
    \label{fig:ad}
\end{figure}

The angular distribution for the 5.505-MeV state is shown in Fig.~\ref{fig:ad}(c). Within the $0f-1p$ orbital space, the $5^-$ state is only accessible through an $\ell=3$, $0f_{7/2}$, neutron transfer on the $^{31}$Si $3/2^+$ ground state. The fit of the calculated DWBA cross sections assuming a pure $\ell=3$ neutron transfer is shown by the lines in Fig.~\ref{fig:ad}(c). The angular distribution data does not show any features indicative of an $\ell = 1$, or any other $\ell<3$, neutron transfer. An upper limit on an $\ell=1$ contribution to the distribution was set at $<10$\% the value of the $\ell=3$ (Table~\ref{tab:si32c2s}). Though $\ell=1$ transfer is not allowed in the 3/2$^+\rightarrow5^-$ transfer, this information puts an upper limit on any other contributing nearby or degenerate states.

Angular distributions for the 4.231-MeV, 5.288-MeV, and 5.772-MeV states are shown in Figs.~\ref{fig:ad}a), (b) and (d) along with the DWBA fits. A corresponding list of (relative) $C^2S$ are listed in Table~\ref{tab:si32c2s}, normalized to the 5.505~MeV state value. For the 5.288-MeV state, the $C^2S$ was reduced by as much as 20\% under the assumption of an additional level at $E_x\approx5.1$~MeV and an upper limit on the $\ell=1$ contribution was extracted. Only a pure $\ell=2$ distribution was used for the 4.231-MeV state as it should be the dominant transfer, however, $\ell=0$ contributions are also allowed. Above $E_x\approx6$~MeV, angular distributions were limited in angle coverage and had no distinct features to signify their $\ell$ values. For completeness, an upper limit on the $C^2S$ of the $E_x=6.347$~MeV level was determined assuming the yield belonged solely to the $4^{(-)}_1$ previously observed state~\cite{ref:Williams2025,ref:Williams2025b}.

\section{Discussion}\label{sec:discussion}

\subsection{\boldmath The $\nu0f_{7/2}$ configuration in the $5^-$ isomeric state}

As laid out in Section~\ref{sec:intro}, the $(0d_{3/2})^1(0f_{7/2})^1$ single-neutron configuration should dominate the wave function of the $5^-$ isomeric level in $^{32}$Si. This independent-particle coupling picture was supported by the energy behaviors of the $5^-$ levels in the even-$A$ systems relative to their corresponding $N=17$ and 19 single-neutron energy differences (Fig.~\ref{fig:sys}). The current single-neutron transfer data establishes the dominant $\ell=3$ neutron transfer to the $5^-$ state. The upper limit on the $\ell=1$ neutron transfer strength is small ($<10$\%) dispelling any significant contribution from the $1p_{3/2}$ neutron orbital in this $E_x$ region [Fig.~\ref{fig:ad}(c)]. Furthermore, the extracted $C^2S$ is relatively large in reference to those of the $2^+_2$ and $3^-_1$ levels (Table~\ref{tab:si32c2s}). The combined $\ell=1+3$ $C^2S$ of the 3$^{(-)}_2$ state at $E_x=5.772$~MeV does have a comparable total. However, as further discussed in sub-Section~\ref{subsec:overlaps}, it is probable that additional states contribute to the cross section around this excitation energy region. Therefore, the single-neutron transfer data confirms a dominant $(0d_{3/2})^1(0f_{7/2})^1$ single-neutron configuration for the $^{32}$Si $5^-$ state's wave function.

\subsection{\boldmath The role of the $\nu0f_{7/2}$ occupancy in the reduced B($E2, 5^-\rightarrow 3^-$) transition strength}

The $\ell=3$ single-neutron overlap to the 5.288-MeV $3^-_1$ level is about 40\% [0.44(11)] of the overlap to the isomeric $5^-_1$ state (Table~\ref{tab:si32c2s}). In the independent-particle picture, this state belongs to the $(0d_{3/2})^1(0f_{7/2})^1$ multiplet which also includes the $5^-_1$ state. However, its reduced $C^2S$ is consistent with the qualitative expectation of a mixed-configuration wave function, as discussed in Section~\ref{sec:background}. 
 The 20\% limit on $\ell=1$ contributions to the $3^-_1$ $C^2S$ leaves room for some $\nu1p_{3/2}$ configurations which is plausible via the $(0d_{3/2})^1(1p_{3/2})^1$ neutron configuration. As discussed in Section~\ref{sec:res}, an additional state present in the $E_x\approx5.1-5.2$~MeV region would reduce the overall $3^-$ strength by as much as 20\%.

In $^{34}$S, the same $3^-_1$ to $5^-_1$ $C^2S$ ratio is $\approx0.25$ showing that the $3^-_1$ has about 20\% the $\ell=3$ overlap strength as the $5^-_1$. This data was obtained from a single-neutron ($d$,$p$) transfer reaction on the $^{33}$S ground state~\cite{ref:Crozier1972,ref:VanDerBaan1971}. The comparable ratio of the $3^-_1$ to $5^-_1$ single-neutron strength between $^{32}$Si [0.44(11)] and $^{34}$S [0.25(4)] is interpreted as similar relative single-neutron $0f_{7/2}$ occupancies between their corresponding states. Under this premise, that the $0f_{7/2}$ occupancies are of the same order for both systems, the orthogonality of the neutron wave functions cannot be the sole 
attribute resulting in the approximately $\times15$ change in their B($E2, 5^-_1\rightarrow 3^-_1$) transition rates.

The hindered $B(E2; 5^-_1\rightarrow 3^-_1)$ transition strength in $^{32}$Si is likely due to the reduced participation of both the neutrons \emph{and} protons in the transition. 
As discussed in Section~\ref{sec:background}, the robust nature of the $Z=14$ sub-shell gap in the $^{32}$Si ground state is linked to the relatively reduced collectivity in the $2^+_1$ state --- a proton distribution that is weakly deformed [Fig.~\ref{fig:sys}(c)] and prolate in shape ($Q_{s}(2^+_1) = 0.11(10)$~$e$b)~\cite{ref:Ibbotson1998,ref:Pro72,ref:Heery2024,ref:Williams2023,ref:Williams2025,ref:Williams2025b}. 
Additionally, the sub-shell produces a relative increase in the $4^+_1$ excitation energy [Fig.~\ref{fig:sys}(a)] due to the required re-arrangement of protons outside the filled $0d_{5/2}$ orbital~\cite{ref:Watwood2025}. 
Despite its $Z=14$ sub-shell, the $B(E2; 2^+_1\rightarrow 0^+_1)$ transition strength in $^{32}$Si is well above $\approx1$~W.u. and is only reduced \emph{relative} to transition strengths in $^{34}$S ($Z=16$) and $^{36}$Ar ($Z=18$) by at most 60\% [Fig.~\ref{fig:sys}(c)]. 
Hence, even if the $^{32}$Si ground-state proton configuration persists in the $5^-_1$ and $3^-_1$ wave functions, it is unlikely that the sphericity of the proton shell closure alone leads to the order-of-magnitude reduction in the transition strength.

In the description of a transition strength in terms of valence single-particle transitions outside a polarizable core, the single-particle matrix elements, $M_{p(n)}$, and the effective charges, $e_{p(n)}$, are related to the transition strength by $B(E2) \propto |e_pM_p + e_nM_n|^2$~\cite{ref:Glaudemans1971,ref:Bohr1975}. 
Therefore, it is proposed that there is a lack of participation by both of the nucleons in the transition which results in the hindrance of the transition strength in $^{32}$Si. 
In $^{32}$Si, a small $M_n$ would follow from the close matching and lack of re-configuration of the $(0d_{3/2})^1(0f_{7/2})^1$ neutron configurations between the two states. 
Considering the relative context of these neutron configurations in two $^{34}$S states, as was discussed above, this alone would not be enough to produce the hindrance. 
This would imply that the single-proton configurations similarly do not differ significantly between the two states and that the $M_p$ in $^{32}$Si is also small. 
Analogous to the neutron case, because the $B(E2; 5^-_1\rightarrow 3^-_1)\approx 1$~W.u. in nearby $^{30}$Si~\cite{ref:Greene1972}, the proton behavior alone could not produce the hindrance. 
Namely, both $^{30}$Si and $^{32}$Si have evidence for robust $Z=14$ sub-shell spacings~\cite{ref:Watwood2025}, however, they differ in the dominant neutron configurations of the $3^-_1$ levels, with the former having a dominant $(1s_{1/2})^1(0f_{7/2})^1$ neutron configuration~\cite{ref:Mackh1973,ref:Baxter1970} as opposed to $(0d_{3/2})^1(0f_{7/2})^1$. 
In addition to the discussion above, different signs for the $M_p$ and $M_n$ could lead to a further reduction in the total transition strength. 
Finally, we note that isoscalar ($M_0$) and isovector ($M_1$) single-particle transition matrix elements must also be reduced and similar in magnitude for small $M_{p(n)}$, as $M_p \propto M_0 + M_1$ and $M_n \propto M_0 - M_1$~\cite{ref:Bernstein1979}.

The $^{30}$Si, $^{32}$Si and $^{34}$Si ground-state proton configurations are each consistent with a robust $Z=14$ shell spacing (See Ref.~\cite{ref:Watwood2025} and references therein). 
For $^{32}$Si and $^{34}$Si, the lowest-lying $3^-$ and $5^-$ levels are also connected through complementary single-neutron configurations involving the $\nu0d_{3/2}$ and $\nu0f_{7/2}$ orbitals, swapping only a particle for a hole between them in the $0d_{3/2}$ orbital. 
This is not the case for the neutron configuration of the lowest $3^-_1$ level in $^{30}$Si which has a large $(1s_{1/2})^1(0f_{7/2})^1$ neutron component~\cite{ref:Mackh1973,ref:Baxter1970}. 
The similar structures of $^{32}$Si and $^{34}$Si suggest that the observation of a very hindered $B(E2; 5^-_1\rightarrow 3^-_1)$ transition in the $N=20$ nucleus $^{34}$Si is probable~\cite{ref:Mutschler2017}. 
In fact, experimental work on the $\gamma$-ray spectrum of $^{34}$Si has yet to observe a connecting transition from the $5^-_1$ state ($E_x = 4.970$~MeV) to the $3^-_1$ level ($E_x = 4.255$~MeV)~\cite{ref:Baumann1989,ref:Nummela2001,ref:Paschalis2011,ref:Rotaru2012,ref:Lica2019,ref:Lubna2024}. 
This is consistent with a hindered $E2$ transition, however, the lifetime of the $5^-_1$ level has yet to be determined and there is an intermediate $4^-_1$ level at $E_x = 4.379$~MeV. 
In the neighboring $N=20$ isotones, $^{36}$S and $^{38}$Ar, the hindrance should be reduced and indeed both nuclei show $5^-_1 \rightarrow 3^-_1$ linking transitions in their excited-level decay schemes~\cite{ref:Grocutt2022,ref:Rudolph2002}. 
In $^{38}$Ar the extracted transition strength, $B(E2; 5^-_1\rightarrow 3^-_1) = 0.223(20)$~W.u.~\cite{ref:Engelbertink1970,ref:Kolata1976}, is similar to the values found in $N=18$ $^{34}$S and $^{36}$Ar [Fig.~\ref{fig:sys}(c)]. 
In the case of $^{36}$S, there is no empirical value for the lifetime of the $5^-_1$ level. 

\subsection{\boldmath Single-neutron overlaps to states above the $5^-$ level}\label{subsec:overlaps}

The multi-fit result of the 5.772-MeV angular distribution produced a summed $C^2S$ value consistent with that of the $5^-$ level [$C^2S$ = 0.91(15)]. It has sizable contributions from both $\ell=1$ and $3$ transfer. In comparison with the $C^2S$ of the $3^-_1$ level, it would be a surprising amount of total strength for a lone $3^{(-)}$ level. An equally probable scenario is that additional states are present in the region. A number of levels are yet to be determined from the $(0d_{3/2})^1(0f_{7/2})^1$ multiplet. For instance, a missing $2^-_1$ state may be a candidate for the small amount of strength observed at $E_x \approx 5.1 - 5.2$~MeV. The $\ell=1$ strength observed must be associated with a member or members of the $J^{\pi} = 0^- - 3^-$, $(0d_{3/2})^1(1p_{3/2})^1$ multiplet of levels and there could be mixing with the lowest $J^\pi = 2^-$ and $3^-$ levels of the $(0d_{3/2})^1(0f_{7/2})^1$ multiplet. The centroid energy differences of the $(0d_{3/2})^1(0f_{7/2})^1$ and the $(0d_{3/2})^1(1p_{3/2})^1$ neutron configurations should be approximately 1~MeV apart~\cite{ref:Chen2024} with the individual members residing even closer in energy as in nearby $^{34}$S for example (see Table 2 of Ref.~\cite{ref:Greene1972}).

The $4^{(-)}$ state suggested at $E_x = 6.347$~MeV~\cite{ref:Williams2023,ref:Williams2025,ref:Williams2025b} is accessible only by $\ell=3$ neutron transfer and a small amount of strength was observed at the correct $E_x$. From the ($t$,$p$) work of Ref.~\cite{ref:For82} a possible $3^-$ level and $1^-$ level could also reside in the region and contribute to the strength. From an independent particle model perspective the expectation would be to observe the $4^-$ of the $(0d_{3/2})^1(0f_{7/2})^1$ multiplet with $\approx 9/11$ the strength of the $5^-$ level. Instead, there are signs of fragmentation of the neutron strength across various $4^-$ levels, or there is a question on the parity of the $J=4$. The former is most probable when considering reference to the single-neutron distribution in $^{34}$S where the strength was found divided between at least the two lowest $4^-$ levels on the order of $\approx2.5:1$ (Table 2 of Ref.~\cite{ref:Crozier1972}).

\section{Summary}\label{sec:sum}

The single-neutron properties of excited levels in $^{32}$Si were characterized via the $^{31}$Si($d$,$p$) reaction. The $5^-$ isomeric state was confirmed to have a dominant $(0d_{3/2})^1(0f_{7/2})^1$ neutron configuration, consistent with the expectations garnered from the cooperative insight of the $N=17-19$ isotones in the region. The relative $\ell=3$ single-neutron spectroscopic strengths of the $3^-_1$ to $5^-$ levels in $^{32}$Si ($\approx0.44$) were found to be comparable to those in $^{34}$S ($\approx0.25$). These results were interpreted as each system having similar relative changes in their single-neutron occupancies between the two states. The consistency of the $^{34}$S and $^{32}$Si single-neutron results contrasts sharply with the order-of-magnitude difference in their B($E2, 5^-\rightarrow 3^-$) transition strengths. Consequently, the hindered $E2$ decay in $^{32}$Si is most probably associated with proton participation, or lack thereof, in the transition. It is of interest whether the same conditions persist in the $N=20$ isotones.

\begin{acknowledgments}
This work was supported by the U.S. Department of Energy, Office of Science, Office of Nuclear Physics, under Contract No. DE-AC02-06CH11357 (Argonne). This research used resources of ANL’s ATLAS facility, which is a DOE Office of Science User Facility. We gratefully acknowledge the computing resources provided on Bebop, a high-performance computing cluster operated by the Laboratory Computing Resource Center at Argonne National Laboratory. 
\end{acknowledgments}

\bibliography{all_references}

\begin{thebibliography}{54}%
\makeatletter
\providecommand \@ifxundefined [1]{%
 \@ifx{#1\undefined}
}%
\providecommand \@ifnum [1]{%
 \ifnum #1\expandafter \@firstoftwo
 \else \expandafter \@secondoftwo
 \fi
}%
\providecommand \@ifx [1]{%
 \ifx #1\expandafter \@firstoftwo
 \else \expandafter \@secondoftwo
 \fi
}%
\providecommand \natexlab [1]{#1}%
\providecommand \enquote  [1]{``#1''}%
\providecommand \bibnamefont  [1]{#1}%
\providecommand \bibfnamefont [1]{#1}%
\providecommand \citenamefont [1]{#1}%
\providecommand \href@noop [0]{\@secondoftwo}%
\providecommand \href [0]{\begingroup \@sanitize@url \@href}%
\providecommand \@href[1]{\@@startlink{#1}\@@href}%
\providecommand \@@href[1]{\endgroup#1\@@endlink}%
\providecommand \@sanitize@url [0]{\catcode `\\12\catcode `\$12\catcode
  `\&12\catcode `\#12\catcode `\^12\catcode `\_12\catcode `\%12\relax}%
\providecommand \@@startlink[1]{}%
\providecommand \@@endlink[0]{}%
\providecommand \url  [0]{\begingroup\@sanitize@url \@url }%
\providecommand \@url [1]{\endgroup\@href {#1}{\urlprefix }}%
\providecommand \urlprefix  [0]{URL }%
\providecommand \Eprint [0]{\href }%
\providecommand \doibase [0]{https://doi.org/}%
\providecommand \selectlanguage [0]{\@gobble}%
\providecommand \bibinfo  [0]{\@secondoftwo}%
\providecommand \bibfield  [0]{\@secondoftwo}%
\providecommand \translation [1]{[#1]}%
\providecommand \BibitemOpen [0]{}%
\providecommand \bibitemStop [0]{}%
\providecommand \bibitemNoStop [0]{.\EOS\space}%
\providecommand \EOS [0]{\spacefactor3000\relax}%
\providecommand \BibitemShut  [1]{\csname bibitem#1\endcsname}%
\let\auto@bib@innerbib\@empty
\bibitem [{\citenamefont {Lindner}(1953)}]{ref:Lindner1953}%
  \BibitemOpen
  \bibfield  {author} {\bibinfo {author} {\bibfnamefont {M.}~\bibnamefont
  {Lindner}},\ }\bibfield  {title} {\bibinfo {title} {New {N}uclides {P}roduced
  in {C}hlorine {S}pallation},\ }\href {https://doi.org/10.1103/PhysRev.91.642}
  {\bibfield  {journal} {\bibinfo  {journal} {Phys. Rev.}\ }\textbf {\bibinfo
  {volume} {91}},\ \bibinfo {pages} {642} (\bibinfo {year} {1953})}\BibitemShut
  {NoStop}%
\bibitem [{\citenamefont {Thoennessen}(2012)}]{ref:Thoennessen2012}%
  \BibitemOpen
  \bibfield  {author} {\bibinfo {author} {\bibfnamefont {M.}~\bibnamefont
  {Thoennessen}},\ }\bibfield  {title} {\bibinfo {title} {Discovery of the
  {I}sotopes with $11\leq{Z}\leq19$},\ }\href
  {https://doi.org/https://doi.org/10.1016/j.adt.2011.09.002} {\bibfield
  {journal} {\bibinfo  {journal} {Atomic Data and Nuclear Data Tables}\
  }\textbf {\bibinfo {volume} {98}},\ \bibinfo {pages} {933} (\bibinfo {year}
  {2012})}\BibitemShut {NoStop}%
\bibitem [{\citenamefont {Fortune}\ \emph {et~al.}(1982)\citenamefont
  {Fortune}, \citenamefont {Bland}, \citenamefont {Watson},\ and\ \citenamefont
  {Abouzeid}}]{ref:For82}%
  \BibitemOpen
  \bibfield  {author} {\bibinfo {author} {\bibfnamefont {H.~T.}\ \bibnamefont
  {Fortune}}, \bibinfo {author} {\bibfnamefont {L.}~\bibnamefont {Bland}},
  \bibinfo {author} {\bibfnamefont {D.~L.}\ \bibnamefont {Watson}},\ and\
  \bibinfo {author} {\bibfnamefont {M.~A.}\ \bibnamefont {Abouzeid}},\
  }\bibfield  {title} {\bibinfo {title} {$^{32}\mathrm{Si}$ from
  $^{30}\mathrm{Si}$($t$,$p$)},\ }\href {https://doi.org/10.1103/PhysRevC.25.5}
  {\bibfield  {journal} {\bibinfo  {journal} {Phys. Rev. C}\ }\textbf {\bibinfo
  {volume} {25}},\ \bibinfo {pages} {5} (\bibinfo {year} {1982})}\BibitemShut
  {NoStop}%
\bibitem [{\citenamefont {Fornal}\ \emph {et~al.}(1997)\citenamefont {Fornal},
  \citenamefont {Broda}, \citenamefont {Kr\'olas}, \citenamefont {Paw\l{}at},
  \citenamefont {Wrzesi\'{n}ski}, \citenamefont {Bazzacco}, \citenamefont
  {Fabris}, \citenamefont {Lunardi}, \citenamefont {Rossi~Alvarez},
  \citenamefont {Viesti}, \citenamefont {de~Angelis}, \citenamefont
  {Cinausero}, \citenamefont {Napoli},\ and\ \citenamefont
  {Grabowski}}]{ref:Fornal1997}%
  \BibitemOpen
  \bibfield  {author} {\bibinfo {author} {\bibfnamefont {B.}~\bibnamefont
  {Fornal}}, \bibinfo {author} {\bibfnamefont {R.}~\bibnamefont {Broda}},
  \bibinfo {author} {\bibfnamefont {W.}~\bibnamefont {Kr\'olas}}, \bibinfo
  {author} {\bibfnamefont {T.}~\bibnamefont {Paw\l{}at}}, \bibinfo {author}
  {\bibfnamefont {J.}~\bibnamefont {Wrzesi\'{n}ski}}, \bibinfo {author}
  {\bibfnamefont {D.}~\bibnamefont {Bazzacco}}, \bibinfo {author}
  {\bibfnamefont {D.}~\bibnamefont {Fabris}}, \bibinfo {author} {\bibfnamefont
  {S.}~\bibnamefont {Lunardi}}, \bibinfo {author} {\bibfnamefont
  {C.}~\bibnamefont {Rossi~Alvarez}}, \bibinfo {author} {\bibfnamefont
  {G.}~\bibnamefont {Viesti}}, \bibinfo {author} {\bibfnamefont
  {G.}~\bibnamefont {de~Angelis}}, \bibinfo {author} {\bibfnamefont
  {M.}~\bibnamefont {Cinausero}}, \bibinfo {author} {\bibfnamefont {D.~R.}\
  \bibnamefont {Napoli}},\ and\ \bibinfo {author} {\bibfnamefont {Z.~W.}\
  \bibnamefont {Grabowski}},\ }\bibfield  {title} {\bibinfo {title}
  {$\ensuremath{\gamma}$-ray studies of neutron-rich $\mathrm{N}=18,19$ nuclei
  produced in deep-inelastic collisions},\ }\href
  {https://doi.org/10.1103/PhysRevC.55.762} {\bibfield  {journal} {\bibinfo
  {journal} {Phys. Rev. C}\ }\textbf {\bibinfo {volume} {55}},\ \bibinfo
  {pages} {762} (\bibinfo {year} {1997})}\BibitemShut {NoStop}%
\bibitem [{\citenamefont {Pronko}\ and\ \citenamefont
  {McDonald}(1972)}]{ref:Pro72}%
  \BibitemOpen
  \bibfield  {author} {\bibinfo {author} {\bibfnamefont {J.~G.}\ \bibnamefont
  {Pronko}}\ and\ \bibinfo {author} {\bibfnamefont {R.~E.}\ \bibnamefont
  {McDonald}},\ }\bibfield  {title} {\bibinfo {title} {Study of
  $^{32}\mathrm{Si}$ using the $^{30}\mathrm{Si}(t,p\ensuremath{\gamma})$
  reaction},\ }\href {https://doi.org/10.1103/PhysRevC.6.2065} {\bibfield
  {journal} {\bibinfo  {journal} {Phys. Rev. C}\ }\textbf {\bibinfo {volume}
  {6}},\ \bibinfo {pages} {2065} (\bibinfo {year} {1972})}\BibitemShut
  {NoStop}%
\bibitem [{\citenamefont {Guillaume}\ \emph {et~al.}(1974)\citenamefont
  {Guillaume}, \citenamefont {Rastegar}, \citenamefont {Fintz},\ and\
  \citenamefont {Gallmann}}]{ref:Gui74}%
  \BibitemOpen
  \bibfield  {author} {\bibinfo {author} {\bibfnamefont {G.}~\bibnamefont
  {Guillaume}}, \bibinfo {author} {\bibfnamefont {B.}~\bibnamefont {Rastegar}},
  \bibinfo {author} {\bibfnamefont {P.}~\bibnamefont {Fintz}},\ and\ \bibinfo
  {author} {\bibfnamefont {A.}~\bibnamefont {Gallmann}},\ }\bibfield  {title}
  {\bibinfo {title} {Transitions \'electromagn\'etiques dans le noyau
  ${}^{32}\mathrm{Si}$ atteint par la r\'eaction ${}^{30}\mathrm{Si}(t,
  p\gamma){}^{32}\mathrm{Si}$},\ }\href
  {https://doi.org/https://doi.org/10.1016/0375-9474(74)90797-0} {\bibfield
  {journal} {\bibinfo  {journal} {Nucl. Phys. A}\ }\textbf {\bibinfo {volume}
  {227}},\ \bibinfo {pages} {284} (\bibinfo {year} {1974})}\BibitemShut
  {NoStop}%
\bibitem [{\citenamefont {Paul}\ \emph {et~al.}(2001)\citenamefont {Paul},
  \citenamefont {R{\"o}ttger}, \citenamefont {Zimbal},\ and\ \citenamefont
  {Keyser}}]{ref:Paul2001}%
  \BibitemOpen
  \bibfield  {author} {\bibinfo {author} {\bibfnamefont {A.}~\bibnamefont
  {Paul}}, \bibinfo {author} {\bibfnamefont {S.}~\bibnamefont {R{\"o}ttger}},
  \bibinfo {author} {\bibfnamefont {A.}~\bibnamefont {Zimbal}},\ and\ \bibinfo
  {author} {\bibfnamefont {U.}~\bibnamefont {Keyser}},\ }\bibfield  {title}
  {\bibinfo {title} {Prompt (n,$\gamma$) mass measurements for the avogadro
  project},\ }\href {https://doi.org/10.1023/A:1011982830022} {\bibfield
  {journal} {\bibinfo  {journal} {Hyperfine Interactions}\ }\textbf {\bibinfo
  {volume} {132}},\ \bibinfo {pages} {189} (\bibinfo {year}
  {2001})}\BibitemShut {NoStop}%
\bibitem [{\citenamefont {Heery}\ \emph {et~al.}(2024)\citenamefont {Heery},
  \citenamefont {Henderson}, \citenamefont {Hoffman}, \citenamefont {Hill},
  \citenamefont {Beck}, \citenamefont {Cousins}, \citenamefont {Farris},
  \citenamefont {Gade}, \citenamefont {Gillespie}, \citenamefont {Holt},
  \citenamefont {Hu}, \citenamefont {Iwasaki}, \citenamefont {Kisyov},
  \citenamefont {Kuchera}, \citenamefont {Longfellow}, \citenamefont
  {M\"uller-Gatermann}, \citenamefont {Poves}, \citenamefont {Rubino},
  \citenamefont {Russell}, \citenamefont {Salinas}, \citenamefont {Sanchez},
  \citenamefont {Weisshaar}, \citenamefont {Wu},\ and\ \citenamefont
  {Wu}}]{ref:Heery2024}%
  \BibitemOpen
  \bibfield  {author} {\bibinfo {author} {\bibfnamefont {J.}~\bibnamefont
  {Heery}}, \bibinfo {author} {\bibfnamefont {J.}~\bibnamefont {Henderson}},
  \bibinfo {author} {\bibfnamefont {C.~R.}\ \bibnamefont {Hoffman}}, \bibinfo
  {author} {\bibfnamefont {A.~M.}\ \bibnamefont {Hill}}, \bibinfo {author}
  {\bibfnamefont {T.}~\bibnamefont {Beck}}, \bibinfo {author} {\bibfnamefont
  {C.}~\bibnamefont {Cousins}}, \bibinfo {author} {\bibfnamefont
  {P.}~\bibnamefont {Farris}}, \bibinfo {author} {\bibfnamefont
  {A.}~\bibnamefont {Gade}}, \bibinfo {author} {\bibfnamefont {S.~A.}\
  \bibnamefont {Gillespie}}, \bibinfo {author} {\bibfnamefont {J.~D.}\
  \bibnamefont {Holt}}, \bibinfo {author} {\bibfnamefont {B.}~\bibnamefont
  {Hu}}, \bibinfo {author} {\bibfnamefont {H.}~\bibnamefont {Iwasaki}},
  \bibinfo {author} {\bibfnamefont {S.}~\bibnamefont {Kisyov}}, \bibinfo
  {author} {\bibfnamefont {A.~N.}\ \bibnamefont {Kuchera}}, \bibinfo {author}
  {\bibfnamefont {B.}~\bibnamefont {Longfellow}}, \bibinfo {author}
  {\bibfnamefont {C.}~\bibnamefont {M\"uller-Gatermann}}, \bibinfo {author}
  {\bibfnamefont {A.}~\bibnamefont {Poves}}, \bibinfo {author} {\bibfnamefont
  {E.}~\bibnamefont {Rubino}}, \bibinfo {author} {\bibfnamefont
  {R.}~\bibnamefont {Russell}}, \bibinfo {author} {\bibfnamefont
  {R.}~\bibnamefont {Salinas}}, \bibinfo {author} {\bibfnamefont
  {A.}~\bibnamefont {Sanchez}}, \bibinfo {author} {\bibfnamefont
  {D.}~\bibnamefont {Weisshaar}}, \bibinfo {author} {\bibfnamefont {C.~Y.}\
  \bibnamefont {Wu}},\ and\ \bibinfo {author} {\bibfnamefont {J.}~\bibnamefont
  {Wu}},\ }\bibfield  {title} {\bibinfo {title} {Suppressed electric quadrupole
  collectivity in $^{32}\mathrm{Si}$},\ }\href
  {https://doi.org/10.1103/PhysRevC.109.014327} {\bibfield  {journal} {\bibinfo
   {journal} {Phys. Rev. C}\ }\textbf {\bibinfo {volume} {109}},\ \bibinfo
  {pages} {014327} (\bibinfo {year} {2024})}\BibitemShut {NoStop}%
\bibitem [{\citenamefont {Williams}\ \emph {et~al.}(2023)\citenamefont
  {Williams}, \citenamefont {Hackman}, \citenamefont {Starosta}, \citenamefont
  {Lubna}, \citenamefont {Choudhary}, \citenamefont {Srivastava}, \citenamefont
  {Andreoiu}, \citenamefont {Annen}, \citenamefont {Asch}, \citenamefont
  {Badanage}, \citenamefont {Ball}, \citenamefont {Beuschlein}, \citenamefont
  {Bidaman}, \citenamefont {Bildstein}, \citenamefont {Coleman}, \citenamefont
  {Garnsworthy}, \citenamefont {Greaves}, \citenamefont {Leckenby},
  \citenamefont {Karayonchev}, \citenamefont {Martin}, \citenamefont {Natzke},
  \citenamefont {Petrache}, \citenamefont {Radich}, \citenamefont
  {Raleigh-Smith}, \citenamefont {Rhodes}, \citenamefont {Russell},
  \citenamefont {Satrazani}, \citenamefont {Spagnoletti}, \citenamefont
  {Svensson}, \citenamefont {Tam}, \citenamefont {Wu}, \citenamefont {Yates},\
  and\ \citenamefont {Yu}}]{ref:Williams2023}%
  \BibitemOpen
  \bibfield  {author} {\bibinfo {author} {\bibfnamefont {J.}~\bibnamefont
  {Williams}}, \bibinfo {author} {\bibfnamefont {G.}~\bibnamefont {Hackman}},
  \bibinfo {author} {\bibfnamefont {K.}~\bibnamefont {Starosta}}, \bibinfo
  {author} {\bibfnamefont {R.~S.}\ \bibnamefont {Lubna}}, \bibinfo {author}
  {\bibfnamefont {P.}~\bibnamefont {Choudhary}}, \bibinfo {author}
  {\bibfnamefont {P.~C.}\ \bibnamefont {Srivastava}}, \bibinfo {author}
  {\bibfnamefont {C.}~\bibnamefont {Andreoiu}}, \bibinfo {author}
  {\bibfnamefont {D.}~\bibnamefont {Annen}}, \bibinfo {author} {\bibfnamefont
  {H.}~\bibnamefont {Asch}}, \bibinfo {author} {\bibfnamefont {M.~D. H. K.~G.}\
  \bibnamefont {Badanage}}, \bibinfo {author} {\bibfnamefont {G.~C.}\
  \bibnamefont {Ball}}, \bibinfo {author} {\bibfnamefont {M.}~\bibnamefont
  {Beuschlein}}, \bibinfo {author} {\bibfnamefont {H.}~\bibnamefont {Bidaman}},
  \bibinfo {author} {\bibfnamefont {V.}~\bibnamefont {Bildstein}}, \bibinfo
  {author} {\bibfnamefont {R.}~\bibnamefont {Coleman}}, \bibinfo {author}
  {\bibfnamefont {A.~B.}\ \bibnamefont {Garnsworthy}}, \bibinfo {author}
  {\bibfnamefont {B.}~\bibnamefont {Greaves}}, \bibinfo {author} {\bibfnamefont
  {G.}~\bibnamefont {Leckenby}}, \bibinfo {author} {\bibfnamefont
  {V.}~\bibnamefont {Karayonchev}}, \bibinfo {author} {\bibfnamefont {M.~S.}\
  \bibnamefont {Martin}}, \bibinfo {author} {\bibfnamefont {C.}~\bibnamefont
  {Natzke}}, \bibinfo {author} {\bibfnamefont {C.~M.}\ \bibnamefont
  {Petrache}}, \bibinfo {author} {\bibfnamefont {A.}~\bibnamefont {Radich}},
  \bibinfo {author} {\bibfnamefont {E.}~\bibnamefont {Raleigh-Smith}}, \bibinfo
  {author} {\bibfnamefont {D.}~\bibnamefont {Rhodes}}, \bibinfo {author}
  {\bibfnamefont {R.}~\bibnamefont {Russell}}, \bibinfo {author} {\bibfnamefont
  {M.}~\bibnamefont {Satrazani}}, \bibinfo {author} {\bibfnamefont
  {P.}~\bibnamefont {Spagnoletti}}, \bibinfo {author} {\bibfnamefont {C.~E.}\
  \bibnamefont {Svensson}}, \bibinfo {author} {\bibfnamefont {D.}~\bibnamefont
  {Tam}}, \bibinfo {author} {\bibfnamefont {F.}~\bibnamefont {Wu}}, \bibinfo
  {author} {\bibfnamefont {D.}~\bibnamefont {Yates}},\ and\ \bibinfo {author}
  {\bibfnamefont {Z.}~\bibnamefont {Yu}},\ }\bibfield  {title} {\bibinfo
  {title} {Identifying the spin-trapped character of the $^{32}\mathrm{Si}$
  isomeric state},\ }\href {https://doi.org/10.1103/PhysRevC.108.L051305}
  {\bibfield  {journal} {\bibinfo  {journal} {Phys. Rev. C}\ }\textbf {\bibinfo
  {volume} {108}},\ \bibinfo {pages} {L051305} (\bibinfo {year}
  {2023})}\BibitemShut {NoStop}%
\bibitem [{\citenamefont {Williams}\ \emph
  {et~al.}(2025{\natexlab{a}})\citenamefont {Williams}, \citenamefont
  {Hackman}, \citenamefont {Starosta}, \citenamefont {Lubna}, \citenamefont
  {Choudhary}, \citenamefont {Sahoo}, \citenamefont {Srivastava}, \citenamefont
  {Andreoiu}, \citenamefont {Annen}, \citenamefont {Asch}, \citenamefont
  {Badanage}, \citenamefont {Ball}, \citenamefont {Beuschlein}, \citenamefont
  {Bidaman}, \citenamefont {Bildstein}, \citenamefont {Coleman}, \citenamefont
  {Garnsworthy}, \citenamefont {Greaves}, \citenamefont {Leckenby},
  \citenamefont {Karayonchev}, \citenamefont {Martin}, \citenamefont {Natzke},
  \citenamefont {Petrache}, \citenamefont {Radich}, \citenamefont
  {Raleigh-Smith}, \citenamefont {Rhodes}, \citenamefont {Russell},
  \citenamefont {Satrazani}, \citenamefont {Spagnoletti}, \citenamefont
  {Svensson}, \citenamefont {Tam}, \citenamefont {Wu}, \citenamefont {Yates},\
  and\ \citenamefont {Yu}}]{ref:Williams2025}%
  \BibitemOpen
  \bibfield  {author} {\bibinfo {author} {\bibfnamefont {J.}~\bibnamefont
  {Williams}}, \bibinfo {author} {\bibfnamefont {G.}~\bibnamefont {Hackman}},
  \bibinfo {author} {\bibfnamefont {K.}~\bibnamefont {Starosta}}, \bibinfo
  {author} {\bibfnamefont {R.~S.}\ \bibnamefont {Lubna}}, \bibinfo {author}
  {\bibfnamefont {P.}~\bibnamefont {Choudhary}}, \bibinfo {author}
  {\bibfnamefont {S.}~\bibnamefont {Sahoo}}, \bibinfo {author} {\bibfnamefont
  {P.~C.}\ \bibnamefont {Srivastava}}, \bibinfo {author} {\bibfnamefont
  {C.}~\bibnamefont {Andreoiu}}, \bibinfo {author} {\bibfnamefont
  {D.}~\bibnamefont {Annen}}, \bibinfo {author} {\bibfnamefont
  {H.}~\bibnamefont {Asch}}, \bibinfo {author} {\bibfnamefont {M.~D. H. K.~G.}\
  \bibnamefont {Badanage}}, \bibinfo {author} {\bibfnamefont {G.~C.}\
  \bibnamefont {Ball}}, \bibinfo {author} {\bibfnamefont {M.}~\bibnamefont
  {Beuschlein}}, \bibinfo {author} {\bibfnamefont {H.}~\bibnamefont {Bidaman}},
  \bibinfo {author} {\bibfnamefont {V.}~\bibnamefont {Bildstein}}, \bibinfo
  {author} {\bibfnamefont {R.}~\bibnamefont {Coleman}}, \bibinfo {author}
  {\bibfnamefont {A.~B.}\ \bibnamefont {Garnsworthy}}, \bibinfo {author}
  {\bibfnamefont {B.}~\bibnamefont {Greaves}}, \bibinfo {author} {\bibfnamefont
  {G.}~\bibnamefont {Leckenby}}, \bibinfo {author} {\bibfnamefont
  {V.}~\bibnamefont {Karayonchev}}, \bibinfo {author} {\bibfnamefont {M.~S.}\
  \bibnamefont {Martin}}, \bibinfo {author} {\bibfnamefont {C.}~\bibnamefont
  {Natzke}}, \bibinfo {author} {\bibfnamefont {C.~M.}\ \bibnamefont
  {Petrache}}, \bibinfo {author} {\bibfnamefont {A.}~\bibnamefont {Radich}},
  \bibinfo {author} {\bibfnamefont {E.}~\bibnamefont {Raleigh-Smith}}, \bibinfo
  {author} {\bibfnamefont {D.}~\bibnamefont {Rhodes}}, \bibinfo {author}
  {\bibfnamefont {R.}~\bibnamefont {Russell}}, \bibinfo {author} {\bibfnamefont
  {M.}~\bibnamefont {Satrazani}}, \bibinfo {author} {\bibfnamefont
  {P.}~\bibnamefont {Spagnoletti}}, \bibinfo {author} {\bibfnamefont {C.~E.}\
  \bibnamefont {Svensson}}, \bibinfo {author} {\bibfnamefont {D.}~\bibnamefont
  {Tam}}, \bibinfo {author} {\bibfnamefont {F.}~\bibnamefont {Wu}}, \bibinfo
  {author} {\bibfnamefont {D.}~\bibnamefont {Yates}},\ and\ \bibinfo {author}
  {\bibfnamefont {Z.}~\bibnamefont {Yu}},\ }\bibfield  {title} {\bibinfo
  {title} {Cross-shell excited states in $^{32}\mathrm{Si}$ and
  $^{29}\mathrm{Al}$ populated using fusion-evaporation},\ }\href
  {https://doi.org/https://doi.org/10.1016/j.nuclphysa.2025.123042} {\bibfield
  {journal} {\bibinfo  {journal} {Nucl. Phys. A}\ }\textbf {\bibinfo {volume}
  {1057}},\ \bibinfo {pages} {123042} (\bibinfo {year}
  {2025}{\natexlab{a}})}\BibitemShut {NoStop}%
\bibitem [{\citenamefont {Williams}\ \emph
  {et~al.}(2025{\natexlab{b}})\citenamefont {Williams}, \citenamefont
  {Hackman}, \citenamefont {Starosta}, \citenamefont {Lubna}, \citenamefont
  {Choudhary}, \citenamefont {Sahoo}, \citenamefont {Srivastava}, \citenamefont
  {Andreoiu}, \citenamefont {Annen}, \citenamefont {Asch}, \citenamefont
  {Badanage}, \citenamefont {Ball}, \citenamefont {Beuschlein}, \citenamefont
  {Bidaman}, \citenamefont {Bildstein}, \citenamefont {Coleman}, \citenamefont
  {Garnsworthy}, \citenamefont {Greaves}, \citenamefont {Leckenby},
  \citenamefont {Karayonchev}, \citenamefont {Martin}, \citenamefont {Natzke},
  \citenamefont {Petrache}, \citenamefont {Radich}, \citenamefont
  {Raleigh-Smith}, \citenamefont {Rhodes}, \citenamefont {Russell},
  \citenamefont {Satrazani}, \citenamefont {Spagnoletti}, \citenamefont
  {Svensson}, \citenamefont {Tam}, \citenamefont {Wu}, \citenamefont {Yates},\
  and\ \citenamefont {Yu}}]{ref:Williams2025b}%
  \BibitemOpen
  \bibfield  {author} {\bibinfo {author} {\bibfnamefont {J.}~\bibnamefont
  {Williams}}, \bibinfo {author} {\bibfnamefont {G.}~\bibnamefont {Hackman}},
  \bibinfo {author} {\bibfnamefont {K.}~\bibnamefont {Starosta}}, \bibinfo
  {author} {\bibfnamefont {R.~S.}\ \bibnamefont {Lubna}}, \bibinfo {author}
  {\bibfnamefont {P.}~\bibnamefont {Choudhary}}, \bibinfo {author}
  {\bibfnamefont {S.}~\bibnamefont {Sahoo}}, \bibinfo {author} {\bibfnamefont
  {P.~C.}\ \bibnamefont {Srivastava}}, \bibinfo {author} {\bibfnamefont
  {C.}~\bibnamefont {Andreoiu}}, \bibinfo {author} {\bibfnamefont
  {D.}~\bibnamefont {Annen}}, \bibinfo {author} {\bibfnamefont
  {H.}~\bibnamefont {Asch}}, \bibinfo {author} {\bibfnamefont {M.~D. H. K.~G.}\
  \bibnamefont {Badanage}}, \bibinfo {author} {\bibfnamefont {G.~C.}\
  \bibnamefont {Ball}}, \bibinfo {author} {\bibfnamefont {M.}~\bibnamefont
  {Beuschlein}}, \bibinfo {author} {\bibfnamefont {H.}~\bibnamefont {Bidaman}},
  \bibinfo {author} {\bibfnamefont {V.}~\bibnamefont {Bildstein}}, \bibinfo
  {author} {\bibfnamefont {R.~J.}\ \bibnamefont {Coleman}}, \bibinfo {author}
  {\bibfnamefont {A.~B.}\ \bibnamefont {Garnsworthy}}, \bibinfo {author}
  {\bibfnamefont {B.}~\bibnamefont {Greaves}}, \bibinfo {author} {\bibfnamefont
  {G.}~\bibnamefont {Leckenby}}, \bibinfo {author} {\bibfnamefont
  {V.}~\bibnamefont {Karayonchev}}, \bibinfo {author} {\bibfnamefont {M.~S.}\
  \bibnamefont {Martin}}, \bibinfo {author} {\bibfnamefont {C.}~\bibnamefont
  {Natzke}}, \bibinfo {author} {\bibfnamefont {C.~M.}\ \bibnamefont
  {Petrache}}, \bibinfo {author} {\bibfnamefont {A.}~\bibnamefont {Radich}},
  \bibinfo {author} {\bibfnamefont {E.}~\bibnamefont {Raleigh-Smith}}, \bibinfo
  {author} {\bibfnamefont {D.}~\bibnamefont {Rhodes}}, \bibinfo {author}
  {\bibfnamefont {R.}~\bibnamefont {Russell}}, \bibinfo {author} {\bibfnamefont
  {M.}~\bibnamefont {Satrazani}}, \bibinfo {author} {\bibfnamefont
  {P.}~\bibnamefont {Spagnoletti}}, \bibinfo {author} {\bibfnamefont {C.~E.}\
  \bibnamefont {Svensson}}, \bibinfo {author} {\bibfnamefont {D.}~\bibnamefont
  {Tam}}, \bibinfo {author} {\bibfnamefont {F.}~\bibnamefont {Wu}}, \bibinfo
  {author} {\bibfnamefont {D.}~\bibnamefont {Yates}},\ and\ \bibinfo {author}
  {\bibfnamefont {Z.}~\bibnamefont {Yu}},\ }\bibfield  {title} {\bibinfo
  {title} {Intruder structures in $^{32}\mathrm{Si}$ and $^{29}\mathrm{Al}$},\
  }\href {https://doi.org/10.1103/2q1s-4517} {\bibfield  {journal} {\bibinfo
  {journal} {Phys. Rev. C}\ }\textbf {\bibinfo {volume} {112}},\ \bibinfo
  {pages} {014318} (\bibinfo {year} {2025}{\natexlab{b}})}\BibitemShut
  {NoStop}%
\bibitem [{\citenamefont {Kib\'edi}\ and\ \citenamefont
  {Spear}(2002)}]{ref:Kibedi2002}%
  \BibitemOpen
  \bibfield  {author} {\bibinfo {author} {\bibfnamefont {T.}~\bibnamefont
  {Kib\'edi}}\ and\ \bibinfo {author} {\bibfnamefont {R.~H.}\ \bibnamefont
  {Spear}},\ }\bibfield  {title} {\bibinfo {title} {Electric octupole ({E}3)
  transition probabilities in nuclei},\ }\href
  {https://doi.org/10.1006/adnd.2001.0871} {\bibfield  {journal} {\bibinfo
  {journal} {Atomic Data and Nuclear Data Tables}\ }\textbf {\bibinfo {volume}
  {80}},\ \bibinfo {pages} {35} (\bibinfo {year} {2002})}\BibitemShut {NoStop}%
\bibitem [{\citenamefont {Greene}\ \emph {et~al.}(1972)\citenamefont {Greene},
  \citenamefont {Kuehner}, \citenamefont {Ball}, \citenamefont {Broude},\ and\
  \citenamefont {Forster}}]{ref:Greene1972}%
  \BibitemOpen
  \bibfield  {author} {\bibinfo {author} {\bibfnamefont {M.~W.}\ \bibnamefont
  {Greene}}, \bibinfo {author} {\bibfnamefont {J.~A.}\ \bibnamefont {Kuehner}},
  \bibinfo {author} {\bibfnamefont {G.~C.}\ \bibnamefont {Ball}}, \bibinfo
  {author} {\bibfnamefont {C.}~\bibnamefont {Broude}},\ and\ \bibinfo {author}
  {\bibfnamefont {J.~S.}\ \bibnamefont {Forster}},\ }\bibfield  {title}
  {\bibinfo {title} {Lifetime of the 5.69 $\mathrm{M}$ev $5^{-}$ level in
  $^{34}\mathrm{S}$},\ }\href {https://doi.org/10.1016/0375-9474(72)90183-2}
  {\bibfield  {journal} {\bibinfo  {journal} {Nucl. Phys. A}\ }\textbf
  {\bibinfo {volume} {188}},\ \bibinfo {pages} {83} (\bibinfo {year}
  {1972})}\BibitemShut {NoStop}%
\bibitem [{\citenamefont {\v{S}. Piskoř}\ \emph {et~al.}(2000)\citenamefont
  {\v{S}. Piskoř}, \citenamefont {Novák}, \citenamefont {Šimečková},
  \citenamefont {Cejpek}, \citenamefont {Kroha}, \citenamefont {Dobeš},\ and\
  \citenamefont {Navrátil}}]{ref:Piskor2000}%
  \BibitemOpen
  \bibfield  {author} {\bibinfo {author} {\bibnamefont {\v{S}. Piskoř}},
  \bibinfo {author} {\bibfnamefont {J.}~\bibnamefont {Novák}}, \bibinfo
  {author} {\bibfnamefont {E.}~\bibnamefont {Šimečková}}, \bibinfo {author}
  {\bibfnamefont {J.}~\bibnamefont {Cejpek}}, \bibinfo {author} {\bibfnamefont
  {V.}~\bibnamefont {Kroha}}, \bibinfo {author} {\bibfnamefont
  {J.}~\bibnamefont {Dobeš}},\ and\ \bibinfo {author} {\bibfnamefont
  {P.}~\bibnamefont {Navrátil}},\ }\bibfield  {title} {\bibinfo {title} {A
  study of the $^{30}${S}i(\textit{d,p})$^{31}${S}i reaction},\ }\href
  {https://doi.org/https://doi.org/10.1016/S0375-9474(99)00425-X} {\bibfield
  {journal} {\bibinfo  {journal} {Nucl. Phys. A}\ }\textbf {\bibinfo {volume}
  {662}},\ \bibinfo {pages} {112} (\bibinfo {year} {2000})}\BibitemShut
  {NoStop}%
\bibitem [{\citenamefont {Crozier}(1972)}]{ref:Crozier1972}%
  \BibitemOpen
  \bibfield  {author} {\bibinfo {author} {\bibfnamefont {D.~J.}\ \bibnamefont
  {Crozier}},\ }\bibfield  {title} {\bibinfo {title} {Energy levels of
  $^{34}\mathrm{S}$ from the $^{33}\mathrm{S}(d$,$p)^{34}\mathrm{S}$
  reaction},\ }\href
  {https://doi.org/https://doi.org/10.1016/0375-9474(72)90780-4} {\bibfield
  {journal} {\bibinfo  {journal} {Nucl. Phys. A}\ }\textbf {\bibinfo {volume}
  {198}},\ \bibinfo {pages} {209} (\bibinfo {year} {1972})}\BibitemShut
  {NoStop}%
\bibitem [{\citenamefont {Van Der~Baan}\ and\ \citenamefont
  {Sikora}(1971)}]{ref:VanDerBaan1971}%
  \BibitemOpen
  \bibfield  {author} {\bibinfo {author} {\bibfnamefont {J.~G.}\ \bibnamefont
  {Van Der~Baan}}\ and\ \bibinfo {author} {\bibfnamefont {B.~R.}\ \bibnamefont
  {Sikora}},\ }\bibfield  {title} {\bibinfo {title} {Investigation of the
  $^{33}\mathrm{S}(d$,$p)^{34}\mathrm{S}$ reaction},\ }\href
  {https://doi.org/https://doi.org/10.1016/0375-9474(71)90963-8} {\bibfield
  {journal} {\bibinfo  {journal} {Nuclear Physics A}\ }\textbf {\bibinfo
  {volume} {173}},\ \bibinfo {pages} {456} (\bibinfo {year}
  {1971})}\BibitemShut {NoStop}%
\bibitem [{\citenamefont {MacGregor}\ \emph {et~al.}(2021)\citenamefont
  {MacGregor}, \citenamefont {Sharp}, \citenamefont {Freeman}, \citenamefont
  {Hoffman}, \citenamefont {Kay}, \citenamefont {Tang}, \citenamefont
  {Gaffney}, \citenamefont {Baader}, \citenamefont {Borge}, \citenamefont
  {Butler}, \citenamefont {Catford}, \citenamefont {Cropper}, \citenamefont
  {de~Angelis}, \citenamefont {Konki}, \citenamefont {Kr\"oll}, \citenamefont
  {Labiche}, \citenamefont {Lazarus}, \citenamefont {Lubna}, \citenamefont
  {Martel}, \citenamefont {McNeel}, \citenamefont {Page}, \citenamefont
  {Poleshchuk}, \citenamefont {Raabe}, \citenamefont {Recchia},\ and\
  \citenamefont {Yang}}]{ref:MacGregor2021}%
  \BibitemOpen
  \bibfield  {author} {\bibinfo {author} {\bibfnamefont {P.~T.}\ \bibnamefont
  {MacGregor}}, \bibinfo {author} {\bibfnamefont {D.~K.}\ \bibnamefont
  {Sharp}}, \bibinfo {author} {\bibfnamefont {S.~J.}\ \bibnamefont {Freeman}},
  \bibinfo {author} {\bibfnamefont {C.~R.}\ \bibnamefont {Hoffman}}, \bibinfo
  {author} {\bibfnamefont {B.~P.}\ \bibnamefont {Kay}}, \bibinfo {author}
  {\bibfnamefont {T.~L.}\ \bibnamefont {Tang}}, \bibinfo {author}
  {\bibfnamefont {L.~P.}\ \bibnamefont {Gaffney}}, \bibinfo {author}
  {\bibfnamefont {E.~F.}\ \bibnamefont {Baader}}, \bibinfo {author}
  {\bibfnamefont {M.~J.~G.}\ \bibnamefont {Borge}}, \bibinfo {author}
  {\bibfnamefont {P.~A.}\ \bibnamefont {Butler}}, \bibinfo {author}
  {\bibfnamefont {W.~N.}\ \bibnamefont {Catford}}, \bibinfo {author}
  {\bibfnamefont {B.~D.}\ \bibnamefont {Cropper}}, \bibinfo {author}
  {\bibfnamefont {G.}~\bibnamefont {de~Angelis}}, \bibinfo {author}
  {\bibfnamefont {J.}~\bibnamefont {Konki}}, \bibinfo {author} {\bibfnamefont
  {T.}~\bibnamefont {Kr\"oll}}, \bibinfo {author} {\bibfnamefont
  {M.}~\bibnamefont {Labiche}}, \bibinfo {author} {\bibfnamefont {I.~H.}\
  \bibnamefont {Lazarus}}, \bibinfo {author} {\bibfnamefont {R.~S.}\
  \bibnamefont {Lubna}}, \bibinfo {author} {\bibfnamefont {I.}~\bibnamefont
  {Martel}}, \bibinfo {author} {\bibfnamefont {D.~G.}\ \bibnamefont {McNeel}},
  \bibinfo {author} {\bibfnamefont {R.~D.}\ \bibnamefont {Page}}, \bibinfo
  {author} {\bibfnamefont {O.}~\bibnamefont {Poleshchuk}}, \bibinfo {author}
  {\bibfnamefont {R.}~\bibnamefont {Raabe}}, \bibinfo {author} {\bibfnamefont
  {F.}~\bibnamefont {Recchia}},\ and\ \bibinfo {author} {\bibfnamefont
  {J.}~\bibnamefont {Yang}},\ }\bibfield  {title} {\bibinfo {title} {Evolution
  of single-particle structure near the $\textit{N}=20$ island of inversion},\
  }\href {https://doi.org/10.1103/PhysRevC.104.L051301} {\bibfield  {journal}
  {\bibinfo  {journal} {Phys. Rev. C}\ }\textbf {\bibinfo {volume} {104}},\
  \bibinfo {pages} {L051301} (\bibinfo {year} {2021})}\BibitemShut {NoStop}%
\bibitem [{\citenamefont {Chen}\ \emph {et~al.}(2024)\citenamefont {Chen},
  \citenamefont {Kay}, \citenamefont {Hoffman}, \citenamefont {Tang},
  \citenamefont {Tolstukhin}, \citenamefont {Bazin}, \citenamefont {Lubna},
  \citenamefont {Ayyad}, \citenamefont {Beceiro-Novo}, \citenamefont {Coombes},
  \citenamefont {Freeman}, \citenamefont {Gaffney}, \citenamefont {Garg},
  \citenamefont {Jayatissa}, \citenamefont {Kuchera}, \citenamefont
  {MacGregor}, \citenamefont {Mitchell}, \citenamefont {Mittig}, \citenamefont
  {Monteagudo}, \citenamefont {Munoz-Ramos}, \citenamefont
  {M{\"u}ller-Gatermann}, \citenamefont {Recchia}, \citenamefont {Rijal},
  \citenamefont {Santamaria}, \citenamefont {Serikow}, \citenamefont {Sharp},
  \citenamefont {Smith}, \citenamefont {Stecenko}, \citenamefont {Wilson},
  \citenamefont {Wuosmaa}, \citenamefont {Yuan}, \citenamefont {Zamora},\ and\
  \citenamefont {Zhang}}]{ref:Chen2024}%
  \BibitemOpen
  \bibfield  {author} {\bibinfo {author} {\bibfnamefont {J.}~\bibnamefont
  {Chen}}, \bibinfo {author} {\bibfnamefont {B.~P.}\ \bibnamefont {Kay}},
  \bibinfo {author} {\bibfnamefont {C.~R.}\ \bibnamefont {Hoffman}}, \bibinfo
  {author} {\bibfnamefont {T.~L.}\ \bibnamefont {Tang}}, \bibinfo {author}
  {\bibfnamefont {I.~A.}\ \bibnamefont {Tolstukhin}}, \bibinfo {author}
  {\bibfnamefont {D.}~\bibnamefont {Bazin}}, \bibinfo {author} {\bibfnamefont
  {R.~S.}\ \bibnamefont {Lubna}}, \bibinfo {author} {\bibfnamefont
  {Y.}~\bibnamefont {Ayyad}}, \bibinfo {author} {\bibfnamefont
  {S.}~\bibnamefont {Beceiro-Novo}}, \bibinfo {author} {\bibfnamefont {B.~J.}\
  \bibnamefont {Coombes}}, \bibinfo {author} {\bibfnamefont {S.~J.}\
  \bibnamefont {Freeman}}, \bibinfo {author} {\bibfnamefont {L.~P.}\
  \bibnamefont {Gaffney}}, \bibinfo {author} {\bibfnamefont {R.}~\bibnamefont
  {Garg}}, \bibinfo {author} {\bibfnamefont {H.}~\bibnamefont {Jayatissa}},
  \bibinfo {author} {\bibfnamefont {A.~N.}\ \bibnamefont {Kuchera}}, \bibinfo
  {author} {\bibfnamefont {P.}~\bibnamefont {MacGregor}}, \bibinfo {author}
  {\bibfnamefont {A.~J.}\ \bibnamefont {Mitchell}}, \bibinfo {author}
  {\bibfnamefont {W.}~\bibnamefont {Mittig}}, \bibinfo {author} {\bibfnamefont
  {B.}~\bibnamefont {Monteagudo}}, \bibinfo {author} {\bibfnamefont
  {A.}~\bibnamefont {Munoz-Ramos}}, \bibinfo {author} {\bibfnamefont
  {C.}~\bibnamefont {M{\"u}ller-Gatermann}}, \bibinfo {author} {\bibfnamefont
  {F.}~\bibnamefont {Recchia}}, \bibinfo {author} {\bibfnamefont
  {N.}~\bibnamefont {Rijal}}, \bibinfo {author} {\bibfnamefont
  {C.}~\bibnamefont {Santamaria}}, \bibinfo {author} {\bibfnamefont {M.~Z.}\
  \bibnamefont {Serikow}}, \bibinfo {author} {\bibfnamefont {D.~K.}\
  \bibnamefont {Sharp}}, \bibinfo {author} {\bibfnamefont {J.}~\bibnamefont
  {Smith}}, \bibinfo {author} {\bibfnamefont {J.~K.}\ \bibnamefont {Stecenko}},
  \bibinfo {author} {\bibfnamefont {G.~L.}\ \bibnamefont {Wilson}}, \bibinfo
  {author} {\bibfnamefont {A.~H.}\ \bibnamefont {Wuosmaa}}, \bibinfo {author}
  {\bibfnamefont {C.~X.}\ \bibnamefont {Yuan}}, \bibinfo {author}
  {\bibfnamefont {J.~C.}\ \bibnamefont {Zamora}},\ and\ \bibinfo {author}
  {\bibfnamefont {Y.~N.}\ \bibnamefont {Zhang}},\ }\bibfield  {title} {\bibinfo
  {title} {Evolution of the nuclear spin-orbit splitting explored via the
  $^{32}\mathrm{Si}$($d$,$p$)$^{33}\mathrm{Si}$ reaction using
  $\mathrm{SOLARIS}$},\ }\href
  {https://doi.org/https://doi.org/10.1016/j.physletb.2024.138678} {\bibfield
  {journal} {\bibinfo  {journal} {Phys. Lett. B}\ }\textbf {\bibinfo {volume}
  {853}},\ \bibinfo {pages} {138678} (\bibinfo {year} {2024})}\BibitemShut
  {NoStop}%
\bibitem [{\citenamefont {Kuchera}\ \emph {et~al.}(2024)\citenamefont
  {Kuchera}, \citenamefont {Hoffman}, \citenamefont {Ryan}, \citenamefont
  {D'Amato}, \citenamefont {Guarinello}, \citenamefont {Kielb}, \citenamefont
  {Aggarwal}, \citenamefont {Ajayi}, \citenamefont {Conley}, \citenamefont
  {Conroy}, \citenamefont {Cottle}, \citenamefont {Esparza}, \citenamefont
  {Genty}, \citenamefont {Hanselman}, \citenamefont {Heinze}, \citenamefont
  {Houlihan}, \citenamefont {Kelly}, \citenamefont {Khawaja}, \citenamefont
  {Lopez-Saavedra}, \citenamefont {McCann}, \citenamefont {Morelock},
  \citenamefont {Riley}, \citenamefont {Sandrik}, \citenamefont {Sitaraman},
  \citenamefont {Spieker}, \citenamefont {Temanson}, \citenamefont {Wibisono},\
  and\ \citenamefont {Wiedenh{\"o}ver}}]{ref:Kuchera2024}%
  \BibitemOpen
  \bibfield  {author} {\bibinfo {author} {\bibfnamefont {A.~N.}\ \bibnamefont
  {Kuchera}}, \bibinfo {author} {\bibfnamefont {C.~R.}\ \bibnamefont
  {Hoffman}}, \bibinfo {author} {\bibfnamefont {G.}~\bibnamefont {Ryan}},
  \bibinfo {author} {\bibfnamefont {I.~B.}\ \bibnamefont {D'Amato}}, \bibinfo
  {author} {\bibfnamefont {O.~M.}\ \bibnamefont {Guarinello}}, \bibinfo
  {author} {\bibfnamefont {P.~S.}\ \bibnamefont {Kielb}}, \bibinfo {author}
  {\bibfnamefont {R.}~\bibnamefont {Aggarwal}}, \bibinfo {author}
  {\bibfnamefont {S.}~\bibnamefont {Ajayi}}, \bibinfo {author} {\bibfnamefont
  {A.~L.}\ \bibnamefont {Conley}}, \bibinfo {author} {\bibfnamefont
  {I.}~\bibnamefont {Conroy}}, \bibinfo {author} {\bibfnamefont {P.~D.}\
  \bibnamefont {Cottle}}, \bibinfo {author} {\bibfnamefont {J.~C.}\
  \bibnamefont {Esparza}}, \bibinfo {author} {\bibfnamefont {S.}~\bibnamefont
  {Genty}}, \bibinfo {author} {\bibfnamefont {K.}~\bibnamefont {Hanselman}},
  \bibinfo {author} {\bibfnamefont {M.}~\bibnamefont {Heinze}}, \bibinfo
  {author} {\bibfnamefont {D.}~\bibnamefont {Houlihan}}, \bibinfo {author}
  {\bibfnamefont {B.}~\bibnamefont {Kelly}}, \bibinfo {author} {\bibfnamefont
  {M.~I.}\ \bibnamefont {Khawaja}}, \bibinfo {author} {\bibfnamefont
  {E.}~\bibnamefont {Lopez-Saavedra}}, \bibinfo {author} {\bibfnamefont
  {G.~W.}\ \bibnamefont {McCann}}, \bibinfo {author} {\bibfnamefont {A.~B.}\
  \bibnamefont {Morelock}}, \bibinfo {author} {\bibfnamefont {L.~A.}\
  \bibnamefont {Riley}}, \bibinfo {author} {\bibfnamefont {A.}~\bibnamefont
  {Sandrik}}, \bibinfo {author} {\bibfnamefont {V.}~\bibnamefont {Sitaraman}},
  \bibinfo {author} {\bibfnamefont {M.}~\bibnamefont {Spieker}}, \bibinfo
  {author} {\bibfnamefont {E.}~\bibnamefont {Temanson}}, \bibinfo {author}
  {\bibfnamefont {C.}~\bibnamefont {Wibisono}},\ and\ \bibinfo {author}
  {\bibfnamefont {I.}~\bibnamefont {Wiedenh{\"o}ver}},\ }\bibfield  {title}
  {\bibinfo {title} {Single-neutron adding on $^{34}\mathrm{S}$},\ }\href
  {https://doi.org/10.1140/epja/s10050-024-01399-z} {\bibfield  {journal}
  {\bibinfo  {journal} {Eur. Phys. J. A}\ }\textbf {\bibinfo {volume} {60}},\
  \bibinfo {pages} {176} (\bibinfo {year} {2024})}\BibitemShut {NoStop}%
\bibitem [{\citenamefont {Burgunder}\ \emph {et~al.}(2014)\citenamefont
  {Burgunder}, \citenamefont {Sorlin}, \citenamefont {Nowacki}, \citenamefont
  {Giron}, \citenamefont {Hammache}, \citenamefont {Moukaddam}, \citenamefont
  {de~S\'er\'eville}, \citenamefont {Beaumel}, \citenamefont {C\`aceres},
  \citenamefont {Cl\'ement}, \citenamefont {Duch\^ene}, \citenamefont {Ebran},
  \citenamefont {Fernandez-Dominguez}, \citenamefont {Flavigny}, \citenamefont
  {Franchoo}, \citenamefont {Gibelin}, \citenamefont {Gillibert}, \citenamefont
  {Gr\'evy}, \citenamefont {Guillot}, \citenamefont {Lepailleur}, \citenamefont
  {Matea}, \citenamefont {Matta}, \citenamefont {Nalpas}, \citenamefont
  {Obertelli}, \citenamefont {Otsuka}, \citenamefont {Pancin}, \citenamefont
  {Poves}, \citenamefont {Raabe}, \citenamefont {Scarpaci}, \citenamefont
  {Stefan}, \citenamefont {Stodel}, \citenamefont {Suzuki},\ and\ \citenamefont
  {Thomas}}]{ref:Burgunder2014}%
  \BibitemOpen
  \bibfield  {author} {\bibinfo {author} {\bibfnamefont {G.}~\bibnamefont
  {Burgunder}}, \bibinfo {author} {\bibfnamefont {O.}~\bibnamefont {Sorlin}},
  \bibinfo {author} {\bibfnamefont {F.}~\bibnamefont {Nowacki}}, \bibinfo
  {author} {\bibfnamefont {S.}~\bibnamefont {Giron}}, \bibinfo {author}
  {\bibfnamefont {F.}~\bibnamefont {Hammache}}, \bibinfo {author}
  {\bibfnamefont {M.}~\bibnamefont {Moukaddam}}, \bibinfo {author}
  {\bibfnamefont {N.}~\bibnamefont {de~S\'er\'eville}}, \bibinfo {author}
  {\bibfnamefont {D.}~\bibnamefont {Beaumel}}, \bibinfo {author} {\bibfnamefont
  {L.}~\bibnamefont {C\`aceres}}, \bibinfo {author} {\bibfnamefont
  {E.}~\bibnamefont {Cl\'ement}}, \bibinfo {author} {\bibfnamefont
  {G.}~\bibnamefont {Duch\^ene}}, \bibinfo {author} {\bibfnamefont {J.~P.}\
  \bibnamefont {Ebran}}, \bibinfo {author} {\bibfnamefont {B.}~\bibnamefont
  {Fernandez-Dominguez}}, \bibinfo {author} {\bibfnamefont {F.}~\bibnamefont
  {Flavigny}}, \bibinfo {author} {\bibfnamefont {S.}~\bibnamefont {Franchoo}},
  \bibinfo {author} {\bibfnamefont {J.}~\bibnamefont {Gibelin}}, \bibinfo
  {author} {\bibfnamefont {A.}~\bibnamefont {Gillibert}}, \bibinfo {author}
  {\bibfnamefont {S.}~\bibnamefont {Gr\'evy}}, \bibinfo {author} {\bibfnamefont
  {J.}~\bibnamefont {Guillot}}, \bibinfo {author} {\bibfnamefont
  {A.}~\bibnamefont {Lepailleur}}, \bibinfo {author} {\bibfnamefont
  {I.}~\bibnamefont {Matea}}, \bibinfo {author} {\bibfnamefont
  {A.}~\bibnamefont {Matta}}, \bibinfo {author} {\bibfnamefont
  {L.}~\bibnamefont {Nalpas}}, \bibinfo {author} {\bibfnamefont
  {A.}~\bibnamefont {Obertelli}}, \bibinfo {author} {\bibfnamefont
  {T.}~\bibnamefont {Otsuka}}, \bibinfo {author} {\bibfnamefont
  {J.}~\bibnamefont {Pancin}}, \bibinfo {author} {\bibfnamefont
  {A.}~\bibnamefont {Poves}}, \bibinfo {author} {\bibfnamefont
  {R.}~\bibnamefont {Raabe}}, \bibinfo {author} {\bibfnamefont {J.~A.}\
  \bibnamefont {Scarpaci}}, \bibinfo {author} {\bibfnamefont {I.}~\bibnamefont
  {Stefan}}, \bibinfo {author} {\bibfnamefont {C.}~\bibnamefont {Stodel}},
  \bibinfo {author} {\bibfnamefont {T.}~\bibnamefont {Suzuki}},\ and\ \bibinfo
  {author} {\bibfnamefont {J.~C.}\ \bibnamefont {Thomas}},\ }\bibfield  {title}
  {\bibinfo {title} {Experimental study of the two-body spin-orbit force in
  nuclei},\ }\href {https://doi.org/10.1103/PhysRevLett.112.042502} {\bibfield
  {journal} {\bibinfo  {journal} {Phys. Rev. Lett.}\ }\textbf {\bibinfo
  {volume} {112}},\ \bibinfo {pages} {042502} (\bibinfo {year}
  {2014})}\BibitemShut {NoStop}%
\bibitem [{\citenamefont {Eckle}\ \emph {et~al.}(1989)\citenamefont {Eckle},
  \citenamefont {Kader}, \citenamefont {Clement}, \citenamefont {Eckle},
  \citenamefont {Merz}, \citenamefont {Hertenberger}, \citenamefont {Maier},
  \citenamefont {Schiemenz},\ and\ \citenamefont {Graw}}]{ref:Eckle1989}%
  \BibitemOpen
  \bibfield  {author} {\bibinfo {author} {\bibfnamefont {G.}~\bibnamefont
  {Eckle}}, \bibinfo {author} {\bibfnamefont {H.}~\bibnamefont {Kader}},
  \bibinfo {author} {\bibfnamefont {H.}~\bibnamefont {Clement}}, \bibinfo
  {author} {\bibfnamefont {F.~J.}\ \bibnamefont {Eckle}}, \bibinfo {author}
  {\bibfnamefont {F.}~\bibnamefont {Merz}}, \bibinfo {author} {\bibfnamefont
  {R.}~\bibnamefont {Hertenberger}}, \bibinfo {author} {\bibfnamefont {H.~J.}\
  \bibnamefont {Maier}}, \bibinfo {author} {\bibfnamefont {P.}~\bibnamefont
  {Schiemenz}},\ and\ \bibinfo {author} {\bibfnamefont {G.}~\bibnamefont
  {Graw}},\ }\bibfield  {title} {\bibinfo {title} {A $^{36}\mathrm{S}$($d$,$p$)
  study with high energy resolution},\ }\href
  {https://doi.org/https://doi.org/10.1016/0375-9474(89)90699-4} {\bibfield
  {journal} {\bibinfo  {journal} {Nucl. Phys. A}\ }\textbf {\bibinfo {volume}
  {491}},\ \bibinfo {pages} {205} (\bibinfo {year} {1989})}\BibitemShut
  {NoStop}%
\bibitem [{\citenamefont {Mermaz}\ \emph {et~al.}(1971)\citenamefont {Mermaz},
  \citenamefont {Whitten}, \citenamefont {Champlin}, \citenamefont {Howard},\
  and\ \citenamefont {Bromley}}]{ref:Mermaz1971}%
  \BibitemOpen
  \bibfield  {author} {\bibinfo {author} {\bibfnamefont {M.~C.}\ \bibnamefont
  {Mermaz}}, \bibinfo {author} {\bibfnamefont {C.~A.}\ \bibnamefont {Whitten}},
  \bibinfo {author} {\bibfnamefont {J.~W.}\ \bibnamefont {Champlin}}, \bibinfo
  {author} {\bibfnamefont {A.~J.}\ \bibnamefont {Howard}},\ and\ \bibinfo
  {author} {\bibfnamefont {D.~A.}\ \bibnamefont {Bromley}},\ }\bibfield
  {title} {\bibinfo {title} {Study of the ($d,p$) reaction on
  $^{28}\mathrm{Si}$, $^{32}\mathrm{S}$, and $^{36}\mathrm{Ar}$ at
  ${E}_{d}=18.00$ $\mathrm{MeV}$},\ }\href
  {https://doi.org/10.1103/PhysRevC.4.1778} {\bibfield  {journal} {\bibinfo
  {journal} {Phys. Rev. C}\ }\textbf {\bibinfo {volume} {4}},\ \bibinfo {pages}
  {1778} (\bibinfo {year} {1971})}\BibitemShut {NoStop}%
\bibitem [{\citenamefont {Raman}\ \emph {et~al.}(2001)\citenamefont {Raman},
  \citenamefont {{Nestor Jr.}},\ and\ \citenamefont
  {Tikkanen}}]{ref:Raman2001}%
  \BibitemOpen
  \bibfield  {author} {\bibinfo {author} {\bibfnamefont {S.}~\bibnamefont
  {Raman}}, \bibinfo {author} {\bibfnamefont {C.}~\bibnamefont {{Nestor
  Jr.}}},\ and\ \bibinfo {author} {\bibfnamefont {P.}~\bibnamefont
  {Tikkanen}},\ }\bibfield  {title} {\bibinfo {title} {Transition probability
  from the ground to the first-excited $2^+$ state of even--even nuclides},\
  }\href {https://doi.org/10.1006/adnd.2001.0858} {\bibfield  {journal}
  {\bibinfo  {journal} {At. Data Nucl. Data Tables}\ }\textbf {\bibinfo
  {volume} {78}},\ \bibinfo {pages} {1} (\bibinfo {year} {2001})}\BibitemShut
  {NoStop}%
\bibitem [{\citenamefont {Watwood}\ \emph {et~al.}(2025)\citenamefont
  {Watwood}, \citenamefont {Hoffman}, \citenamefont {Kay}, \citenamefont
  {Tolstukhin}, \citenamefont {Chen}, \citenamefont {Tang}, \citenamefont
  {Bazin}, \citenamefont {Ayyad}, \citenamefont {Beceiro-Novo}, \citenamefont
  {Freeman}, \citenamefont {Gaffney}, \citenamefont {Garg}, \citenamefont
  {Jayatissa}, \citenamefont {Kuchera}, \citenamefont {MacGregor},
  \citenamefont {Mitchell}, \citenamefont {Mu\~noz Ramos}, \citenamefont
  {M\"uller-Gatermann}, \citenamefont {Recchia}, \citenamefont {Santamaria},
  \citenamefont {Serikow}, \citenamefont {Sharp}, \citenamefont {Wilson},
  \citenamefont {Wuosmaa},\ and\ \citenamefont {Zamora}}]{ref:Watwood2025}%
  \BibitemOpen
  \bibfield  {author} {\bibinfo {author} {\bibfnamefont {N.}~\bibnamefont
  {Watwood}}, \bibinfo {author} {\bibfnamefont {C.~R.}\ \bibnamefont
  {Hoffman}}, \bibinfo {author} {\bibfnamefont {B.~P.}\ \bibnamefont {Kay}},
  \bibinfo {author} {\bibfnamefont {I.~A.}\ \bibnamefont {Tolstukhin}},
  \bibinfo {author} {\bibfnamefont {J.}~\bibnamefont {Chen}}, \bibinfo {author}
  {\bibfnamefont {T.~L.}\ \bibnamefont {Tang}}, \bibinfo {author}
  {\bibfnamefont {D.}~\bibnamefont {Bazin}}, \bibinfo {author} {\bibfnamefont
  {Y.}~\bibnamefont {Ayyad}}, \bibinfo {author} {\bibfnamefont
  {S.}~\bibnamefont {Beceiro-Novo}}, \bibinfo {author} {\bibfnamefont {S.~J.}\
  \bibnamefont {Freeman}}, \bibinfo {author} {\bibfnamefont {L.~P.}\
  \bibnamefont {Gaffney}}, \bibinfo {author} {\bibfnamefont {R.}~\bibnamefont
  {Garg}}, \bibinfo {author} {\bibfnamefont {H.}~\bibnamefont {Jayatissa}},
  \bibinfo {author} {\bibfnamefont {A.~N.}\ \bibnamefont {Kuchera}}, \bibinfo
  {author} {\bibfnamefont {P.~T.}\ \bibnamefont {MacGregor}}, \bibinfo {author}
  {\bibfnamefont {A.~J.}\ \bibnamefont {Mitchell}}, \bibinfo {author}
  {\bibfnamefont {A.}~\bibnamefont {Mu\~noz Ramos}}, \bibinfo {author}
  {\bibfnamefont {C.}~\bibnamefont {M\"uller-Gatermann}}, \bibinfo {author}
  {\bibfnamefont {F.}~\bibnamefont {Recchia}}, \bibinfo {author} {\bibfnamefont
  {C.}~\bibnamefont {Santamaria}}, \bibinfo {author} {\bibfnamefont {M.~Z.}\
  \bibnamefont {Serikow}}, \bibinfo {author} {\bibfnamefont {D.~K.}\
  \bibnamefont {Sharp}}, \bibinfo {author} {\bibfnamefont {G.~L.}\ \bibnamefont
  {Wilson}}, \bibinfo {author} {\bibfnamefont {A.~H.}\ \bibnamefont
  {Wuosmaa}},\ and\ \bibinfo {author} {\bibfnamefont {J.~C.}\ \bibnamefont
  {Zamora}},\ }\bibfield  {title} {\bibinfo {title} {Valence
  $1s\text{\ensuremath{-}}0d$ proton vacancy of the $^{32}\mathrm{Si}$ ground
  state},\ }\href {https://doi.org/10.1103/4lzs-mv3l} {\bibfield  {journal}
  {\bibinfo  {journal} {Phys. Rev. C}\ }\textbf {\bibinfo {volume} {112}},\
  \bibinfo {pages} {054304} (\bibinfo {year} {2025})}\BibitemShut {NoStop}%
\bibitem [{\citenamefont {Bohr}\ and\ \citenamefont
  {Mottelson}(1975)}]{ref:Bohr1975}%
  \BibitemOpen
  \bibfield  {author} {\bibinfo {author} {\bibfnamefont {A.}~\bibnamefont
  {Bohr}}\ and\ \bibinfo {author} {\bibfnamefont {B.~R.}\ \bibnamefont
  {Mottelson}},\ }\href@noop {} {\emph {\bibinfo {title} {Nuclear Structure:
  Nuclear Deformations}}},\ Vol.~\bibinfo {volume} {2}\ (\bibinfo  {publisher}
  {Benjamin},\ \bibinfo {address} {Reading, MA},\ \bibinfo {year}
  {1975})\BibitemShut {NoStop}%
\bibitem [{\citenamefont {Spear}(1989)}]{ref:Spear1989}%
  \BibitemOpen
  \bibfield  {author} {\bibinfo {author} {\bibfnamefont {R.~H.}\ \bibnamefont
  {Spear}},\ }\bibfield  {title} {\bibinfo {title} {Reduced electric-octupole
  transition probabilities, {$B(E3; 0^+ \to 3^-_1)$}, for even-even nuclides
  throughout the periodic table},\ }\href
  {https://doi.org/10.1016/0092-640X(89)90008-8} {\bibfield  {journal}
  {\bibinfo  {journal} {Atomic Data and Nuclear Data Tables}\ }\textbf
  {\bibinfo {volume} {42}},\ \bibinfo {pages} {55} (\bibinfo {year}
  {1989})}\BibitemShut {NoStop}%
\bibitem [{\citenamefont {Lubna}\ \emph {et~al.}(2020)\citenamefont {Lubna},
  \citenamefont {Kravvaris}, \citenamefont {Tabor}, \citenamefont {Tripathi},
  \citenamefont {Rubino},\ and\ \citenamefont {Volya}}]{ref:Lubna2020}%
  \BibitemOpen
  \bibfield  {author} {\bibinfo {author} {\bibfnamefont {R.~S.}\ \bibnamefont
  {Lubna}}, \bibinfo {author} {\bibfnamefont {K.}~\bibnamefont {Kravvaris}},
  \bibinfo {author} {\bibfnamefont {S.~L.}\ \bibnamefont {Tabor}}, \bibinfo
  {author} {\bibfnamefont {V.}~\bibnamefont {Tripathi}}, \bibinfo {author}
  {\bibfnamefont {E.}~\bibnamefont {Rubino}},\ and\ \bibinfo {author}
  {\bibfnamefont {A.}~\bibnamefont {Volya}},\ }\bibfield  {title} {\bibinfo
  {title} {Evolution of the $\mathrm{N}=20$ and 28 shell gaps and
  two-particle-two-hole states in the $\mathrm{FSU}$ interaction},\ }\href
  {https://doi.org/10.1103/PhysRevResearch.2.043342} {\bibfield  {journal}
  {\bibinfo  {journal} {Phys. Rev. Research}\ }\textbf {\bibinfo {volume}
  {2}},\ \bibinfo {pages} {043342} (\bibinfo {year} {2020})}\BibitemShut
  {NoStop}%
\bibitem [{\citenamefont {Lubna}\ \emph {et~al.}(2024)\citenamefont {Lubna},
  \citenamefont {Garnsworthy}, \citenamefont {Tripathi}, \citenamefont {Ball},
  \citenamefont {Natzke}, \citenamefont {Rocchini}, \citenamefont {Andreoiu},
  \citenamefont {Bhattacharjee}, \citenamefont {Dillmann}, \citenamefont
  {Garcia}, \citenamefont {Gillespie}, \citenamefont {Hackman}, \citenamefont
  {Griffin}, \citenamefont {Leckenby}, \citenamefont {Miyagi}, \citenamefont
  {Olaizola}, \citenamefont {Porzio}, \citenamefont {Rajabali}, \citenamefont
  {Saito}, \citenamefont {Spagnoletti}, \citenamefont {Tabor}, \citenamefont
  {Umashankar}, \citenamefont {Vedia}, \citenamefont {Volya}, \citenamefont
  {Williams},\ and\ \citenamefont {Yates}}]{ref:Lubna2024}%
  \BibitemOpen
  \bibfield  {author} {\bibinfo {author} {\bibfnamefont {R.~S.}\ \bibnamefont
  {Lubna}}, \bibinfo {author} {\bibfnamefont {A.~B.}\ \bibnamefont
  {Garnsworthy}}, \bibinfo {author} {\bibfnamefont {V.}~\bibnamefont
  {Tripathi}}, \bibinfo {author} {\bibfnamefont {G.~C.}\ \bibnamefont {Ball}},
  \bibinfo {author} {\bibfnamefont {C.~R.}\ \bibnamefont {Natzke}}, \bibinfo
  {author} {\bibfnamefont {M.}~\bibnamefont {Rocchini}}, \bibinfo {author}
  {\bibfnamefont {C.}~\bibnamefont {Andreoiu}}, \bibinfo {author}
  {\bibfnamefont {S.~S.}\ \bibnamefont {Bhattacharjee}}, \bibinfo {author}
  {\bibfnamefont {I.}~\bibnamefont {Dillmann}}, \bibinfo {author}
  {\bibfnamefont {F.~H.}\ \bibnamefont {Garcia}}, \bibinfo {author}
  {\bibfnamefont {S.~A.}\ \bibnamefont {Gillespie}}, \bibinfo {author}
  {\bibfnamefont {G.}~\bibnamefont {Hackman}}, \bibinfo {author} {\bibfnamefont
  {C.~J.}\ \bibnamefont {Griffin}}, \bibinfo {author} {\bibfnamefont
  {G.}~\bibnamefont {Leckenby}}, \bibinfo {author} {\bibfnamefont
  {T.}~\bibnamefont {Miyagi}}, \bibinfo {author} {\bibfnamefont
  {B.}~\bibnamefont {Olaizola}}, \bibinfo {author} {\bibfnamefont
  {C.}~\bibnamefont {Porzio}}, \bibinfo {author} {\bibfnamefont {M.~M.}\
  \bibnamefont {Rajabali}}, \bibinfo {author} {\bibfnamefont {Y.}~\bibnamefont
  {Saito}}, \bibinfo {author} {\bibfnamefont {P.}~\bibnamefont {Spagnoletti}},
  \bibinfo {author} {\bibfnamefont {S.~L.}\ \bibnamefont {Tabor}}, \bibinfo
  {author} {\bibfnamefont {R.}~\bibnamefont {Umashankar}}, \bibinfo {author}
  {\bibfnamefont {V.}~\bibnamefont {Vedia}}, \bibinfo {author} {\bibfnamefont
  {A.}~\bibnamefont {Volya}}, \bibinfo {author} {\bibfnamefont
  {J.}~\bibnamefont {Williams}},\ and\ \bibinfo {author} {\bibfnamefont
  {D.}~\bibnamefont {Yates}},\ }\bibfield  {title} {\bibinfo {title}
  {Cross-shell excited configurations in the structure of $^{34}\mathrm{Si}$},\
  }\href {https://doi.org/10.1103/PhysRevC.109.014309} {\bibfield  {journal}
  {\bibinfo  {journal} {Phys. Rev. C}\ }\textbf {\bibinfo {volume} {109}},\
  \bibinfo {pages} {014309} (\bibinfo {year} {2024})}\BibitemShut {NoStop}%
\bibitem [{\citenamefont {Hoffman}\ \emph {et~al.}(2022)\citenamefont
  {Hoffman}, \citenamefont {Tang}, \citenamefont {Avila}, \citenamefont
  {Ayyad}, \citenamefont {Brown}, \citenamefont {Chen}, \citenamefont {Chipps},
  \citenamefont {Jayatissa}, \citenamefont {Kay}, \citenamefont
  {M\"uller-Gatermann}, \citenamefont {Ong}, \citenamefont {Song},\ and\
  \citenamefont {Wilson}}]{ref:Hoffman2022}%
  \BibitemOpen
  \bibfield  {author} {\bibinfo {author} {\bibfnamefont {C.}~\bibnamefont
  {Hoffman}}, \bibinfo {author} {\bibfnamefont {T.}~\bibnamefont {Tang}},
  \bibinfo {author} {\bibfnamefont {M.}~\bibnamefont {Avila}}, \bibinfo
  {author} {\bibfnamefont {Y.}~\bibnamefont {Ayyad}}, \bibinfo {author}
  {\bibfnamefont {K.}~\bibnamefont {Brown}}, \bibinfo {author} {\bibfnamefont
  {J.}~\bibnamefont {Chen}}, \bibinfo {author} {\bibfnamefont {K.}~\bibnamefont
  {Chipps}}, \bibinfo {author} {\bibfnamefont {H.}~\bibnamefont {Jayatissa}},
  \bibinfo {author} {\bibfnamefont {B.}~\bibnamefont {Kay}}, \bibinfo {author}
  {\bibfnamefont {C.}~\bibnamefont {M\"uller-Gatermann}}, \bibinfo {author}
  {\bibfnamefont {H.}~\bibnamefont {Ong}}, \bibinfo {author} {\bibfnamefont
  {J.}~\bibnamefont {Song}},\ and\ \bibinfo {author} {\bibfnamefont
  {G.}~\bibnamefont {Wilson}},\ }\bibfield  {title} {\bibinfo {title}
  {In-flight production of an isomeric beam of $^{16}${N}},\ }\href
  {https://doi.org/https://doi.org/10.1016/j.nima.2022.166612} {\bibfield
  {journal} {\bibinfo  {journal} {Nucl. Instrum. Meth. A}\ }\textbf {\bibinfo
  {volume} {1032}},\ \bibinfo {pages} {166612} (\bibinfo {year}
  {2022})}\BibitemShut {NoStop}%
\bibitem [{\citenamefont {Rehm}\ \emph {et~al.}(2011)\citenamefont {Rehm},
  \citenamefont {Greene}, \citenamefont {Harss}, \citenamefont {Henderson},
  \citenamefont {Jiang}, \citenamefont {Pardo}, \citenamefont {Zabransky},\
  and\ \citenamefont {Paul}}]{ref:Rehm2011}%
  \BibitemOpen
  \bibfield  {author} {\bibinfo {author} {\bibfnamefont {K.}~\bibnamefont
  {Rehm}}, \bibinfo {author} {\bibfnamefont {J.}~\bibnamefont {Greene}},
  \bibinfo {author} {\bibfnamefont {B.}~\bibnamefont {Harss}}, \bibinfo
  {author} {\bibfnamefont {D.}~\bibnamefont {Henderson}}, \bibinfo {author}
  {\bibfnamefont {C.}~\bibnamefont {Jiang}}, \bibinfo {author} {\bibfnamefont
  {R.}~\bibnamefont {Pardo}}, \bibinfo {author} {\bibfnamefont
  {B.}~\bibnamefont {Zabransky}},\ and\ \bibinfo {author} {\bibfnamefont
  {M.}~\bibnamefont {Paul}},\ }\bibfield  {title} {\bibinfo {title} {Gas cell
  targets for experiments with radioactive beams},\ }\href
  {https://doi.org/https://doi.org/10.1016/j.nima.2011.04.011} {\bibfield
  {journal} {\bibinfo  {journal} {Nucl. Instrum. Meth. A}\ }\textbf {\bibinfo
  {volume} {647}},\ \bibinfo {pages} {3} (\bibinfo {year} {2011})}\BibitemShut
  {NoStop}%
\bibitem [{\citenamefont {Wuosmaa}\ \emph {et~al.}(2007)\citenamefont
  {Wuosmaa}, \citenamefont {Schiffer}, \citenamefont {Back}, \citenamefont
  {Lister},\ and\ \citenamefont {Rehm}}]{ref:Wuo07}%
  \BibitemOpen
  \bibfield  {author} {\bibinfo {author} {\bibfnamefont {A.~H.}\ \bibnamefont
  {Wuosmaa}}, \bibinfo {author} {\bibfnamefont {J.~P.}\ \bibnamefont
  {Schiffer}}, \bibinfo {author} {\bibfnamefont {B.~B.}\ \bibnamefont {Back}},
  \bibinfo {author} {\bibfnamefont {C.~J.}\ \bibnamefont {Lister}},\ and\
  \bibinfo {author} {\bibfnamefont {K.~E.}\ \bibnamefont {Rehm}},\ }\bibfield
  {title} {\bibinfo {title} {A solenoidal spectrometer for reactions in inverse
  kinematics},\ }\href
  {https://www.sciencedirect.com/science/article/abs/pii/S0168900207014490}
  {\bibfield  {journal} {\bibinfo  {journal} {Nucl. Instrum. Meth. A}\ }\textbf
  {\bibinfo {volume} {580}},\ \bibinfo {pages} {1290 } (\bibinfo {year}
  {2007})}\BibitemShut {NoStop}%
\bibitem [{\citenamefont {Lighthall}\ \emph {et~al.}(2010)\citenamefont
  {Lighthall}, \citenamefont {Back}, \citenamefont {Baker}, \citenamefont
  {Freeman}, \citenamefont {Lee}, \citenamefont {Kay}, \citenamefont {Marley},
  \citenamefont {Rehm}, \citenamefont {Rohrer}, \citenamefont {Schiffer},
  \citenamefont {Shetty}, \citenamefont {Vann}, \citenamefont {Winkelbauer},\
  and\ \citenamefont {Wuosmaa}}]{ref:Lig10}%
  \BibitemOpen
  \bibfield  {author} {\bibinfo {author} {\bibfnamefont {J.~C.}\ \bibnamefont
  {Lighthall}}, \bibinfo {author} {\bibfnamefont {B.~B.}\ \bibnamefont {Back}},
  \bibinfo {author} {\bibfnamefont {S.~I.}\ \bibnamefont {Baker}}, \bibinfo
  {author} {\bibfnamefont {S.~J.}\ \bibnamefont {Freeman}}, \bibinfo {author}
  {\bibfnamefont {H.~Y.}\ \bibnamefont {Lee}}, \bibinfo {author} {\bibfnamefont
  {B.~P.}\ \bibnamefont {Kay}}, \bibinfo {author} {\bibfnamefont {S.~T.}\
  \bibnamefont {Marley}}, \bibinfo {author} {\bibfnamefont {K.~E.}\
  \bibnamefont {Rehm}}, \bibinfo {author} {\bibfnamefont {J.~E.}\ \bibnamefont
  {Rohrer}}, \bibinfo {author} {\bibfnamefont {J.~P.}\ \bibnamefont
  {Schiffer}}, \bibinfo {author} {\bibfnamefont {D.~V.}\ \bibnamefont
  {Shetty}}, \bibinfo {author} {\bibfnamefont {A.~W.}\ \bibnamefont {Vann}},
  \bibinfo {author} {\bibfnamefont {J.~R.}\ \bibnamefont {Winkelbauer}},\ and\
  \bibinfo {author} {\bibfnamefont {A.~H.}\ \bibnamefont {Wuosmaa}},\
  }\bibfield  {title} {\bibinfo {title} {Commissioning of the $\mathrm{HELIOS}$
  spectrometer},\ }\href
  {https://doi.org/https://doi.org/10.1016/j.nima.2010.06.220} {\bibfield
  {journal} {\bibinfo  {journal} {Nucl. Instrum. Meth. A}\ }\textbf {\bibinfo
  {volume} {622}},\ \bibinfo {pages} {97} (\bibinfo {year} {2010})}\BibitemShut
  {NoStop}%
\bibitem [{\citenamefont {Hoffman}\ \emph {et~al.}(2012)\citenamefont
  {Hoffman}, \citenamefont {Back}, \citenamefont {Kay}, \citenamefont
  {Schiffer}, \citenamefont {Alcorta}, \citenamefont {Baker}, \citenamefont
  {Bedoor}, \citenamefont {Bertone}, \citenamefont {Clark}, \citenamefont
  {Deibel}, \citenamefont {DiGiovine}, \citenamefont {Freeman}, \citenamefont
  {Greene}, \citenamefont {Lighthall}, \citenamefont {Marley}, \citenamefont
  {Pardo}, \citenamefont {Rehm}, \citenamefont {Rojas}, \citenamefont
  {Santiago-Gonzalez}, \citenamefont {Sharp}, \citenamefont {Shetty},
  \citenamefont {Thomas}, \citenamefont {Wiedenh\"over},\ and\ \citenamefont
  {Wuosmaa}}]{ref:Hoffman2012}%
  \BibitemOpen
  \bibfield  {author} {\bibinfo {author} {\bibfnamefont {C.~R.}\ \bibnamefont
  {Hoffman}}, \bibinfo {author} {\bibfnamefont {B.~B.}\ \bibnamefont {Back}},
  \bibinfo {author} {\bibfnamefont {B.~P.}\ \bibnamefont {Kay}}, \bibinfo
  {author} {\bibfnamefont {J.~P.}\ \bibnamefont {Schiffer}}, \bibinfo {author}
  {\bibfnamefont {M.}~\bibnamefont {Alcorta}}, \bibinfo {author} {\bibfnamefont
  {S.~I.}\ \bibnamefont {Baker}}, \bibinfo {author} {\bibfnamefont
  {S.}~\bibnamefont {Bedoor}}, \bibinfo {author} {\bibfnamefont {P.~F.}\
  \bibnamefont {Bertone}}, \bibinfo {author} {\bibfnamefont {J.~A.}\
  \bibnamefont {Clark}}, \bibinfo {author} {\bibfnamefont {C.~M.}\ \bibnamefont
  {Deibel}}, \bibinfo {author} {\bibfnamefont {B.}~\bibnamefont {DiGiovine}},
  \bibinfo {author} {\bibfnamefont {S.~J.}\ \bibnamefont {Freeman}}, \bibinfo
  {author} {\bibfnamefont {J.~P.}\ \bibnamefont {Greene}}, \bibinfo {author}
  {\bibfnamefont {J.~C.}\ \bibnamefont {Lighthall}}, \bibinfo {author}
  {\bibfnamefont {S.~T.}\ \bibnamefont {Marley}}, \bibinfo {author}
  {\bibfnamefont {R.~C.}\ \bibnamefont {Pardo}}, \bibinfo {author}
  {\bibfnamefont {K.~E.}\ \bibnamefont {Rehm}}, \bibinfo {author}
  {\bibfnamefont {A.}~\bibnamefont {Rojas}}, \bibinfo {author} {\bibfnamefont
  {D.}~\bibnamefont {Santiago-Gonzalez}}, \bibinfo {author} {\bibfnamefont
  {D.~K.}\ \bibnamefont {Sharp}}, \bibinfo {author} {\bibfnamefont {D.~V.}\
  \bibnamefont {Shetty}}, \bibinfo {author} {\bibfnamefont {J.~S.}\
  \bibnamefont {Thomas}}, \bibinfo {author} {\bibfnamefont {I.}~\bibnamefont
  {Wiedenh\"over}},\ and\ \bibinfo {author} {\bibfnamefont {A.~H.}\
  \bibnamefont {Wuosmaa}},\ }\bibfield  {title} {\bibinfo {title} {Experimental
  study of the $^{19}\textrm{O}(d$,$p)^{20}\textrm{O}$ reaction in inverse
  kinematics},\ }\href {https://doi.org/10.1103/PhysRevC.85.054318} {\bibfield
  {journal} {\bibinfo  {journal} {Phys. Rev. C}\ }\textbf {\bibinfo {volume}
  {85}},\ \bibinfo {pages} {054318} (\bibinfo {year} {2012})}\BibitemShut
  {NoStop}%
\bibitem [{\citenamefont {Macfarlane}\ and\ \citenamefont
  {Pieper}(1978)}]{ref:Mac78}%
  \BibitemOpen
  \bibfield  {author} {\bibinfo {author} {\bibfnamefont {M.~H.}\ \bibnamefont
  {Macfarlane}}\ and\ \bibinfo {author} {\bibfnamefont {S.~C.}\ \bibnamefont
  {Pieper}},\ }\href@noop {} {\bibinfo {title} {Ptolemy: A program for
  heavy-ion direct-reaction calculations}} (\bibinfo {year} {1978}),\ \bibinfo
  {note} {{A}rgonne {N}ational {L}aboratory {R}eport ANL-76-11, Rev.
  1}\BibitemShut {NoStop}%
\bibitem [{\citenamefont {An}\ and\ \citenamefont {Cai}(2006)}]{ref:An06}%
  \BibitemOpen
  \bibfield  {author} {\bibinfo {author} {\bibfnamefont {H.}~\bibnamefont
  {An}}\ and\ \bibinfo {author} {\bibfnamefont {C.}~\bibnamefont {Cai}},\
  }\bibfield  {title} {\bibinfo {title} {Global deuteron optical model
  potential for the energy range up to 183 $\mathrm{MeV}$},\ }\href
  {https://doi.org/10.1103/PhysRevC.73.054605} {\bibfield  {journal} {\bibinfo
  {journal} {Phys. Rev. C}\ }\textbf {\bibinfo {volume} {73}},\ \bibinfo
  {pages} {054605} (\bibinfo {year} {2006})}\BibitemShut {NoStop}%
\bibitem [{\citenamefont {Koning}\ and\ \citenamefont
  {Delaroche}(2003)}]{ref:Koning2003}%
  \BibitemOpen
  \bibfield  {author} {\bibinfo {author} {\bibfnamefont {A.~J.}\ \bibnamefont
  {Koning}}\ and\ \bibinfo {author} {\bibfnamefont {J.~P.}\ \bibnamefont
  {Delaroche}},\ }\bibfield  {title} {\bibinfo {title} {Local and global
  nucleon optical models from 1 ke$\mathrm{V}$ to 200
  $\mathrm{M}$e$\mathrm{V}$},\ }\href
  {https://doi.org/https://doi.org/10.1016/S0375-9474(02)01321-0} {\bibfield
  {journal} {\bibinfo  {journal} {Nucl. Phys. A}\ }\textbf {\bibinfo {volume}
  {713}},\ \bibinfo {pages} {231} (\bibinfo {year} {2003})}\BibitemShut
  {NoStop}%
\bibitem [{\citenamefont {Perey}\ and\ \citenamefont
  {Perey}(1963)}]{ref:Perey1963d}%
  \BibitemOpen
  \bibfield  {author} {\bibinfo {author} {\bibfnamefont {C.~M.}\ \bibnamefont
  {Perey}}\ and\ \bibinfo {author} {\bibfnamefont {F.~G.}\ \bibnamefont
  {Perey}},\ }\bibfield  {title} {\bibinfo {title} {Deuteron optical-model
  analysis in the range of 11 to 27 $\textrm{MeV}$},\ }\href
  {https://doi.org/10.1103/PhysRev.132.755} {\bibfield  {journal} {\bibinfo
  {journal} {Phys. Rev.}\ }\textbf {\bibinfo {volume} {132}},\ \bibinfo {pages}
  {755} (\bibinfo {year} {1963})}\BibitemShut {NoStop}%
\bibitem [{\citenamefont {Perey}(1963)}]{ref:Perey1963p}%
  \BibitemOpen
  \bibfield  {author} {\bibinfo {author} {\bibfnamefont {F.~G.}\ \bibnamefont
  {Perey}},\ }\bibfield  {title} {\bibinfo {title} {Optical-model analysis of
  proton elastic scattering in the range of 9 to 22 $\textrm{MeV}$},\ }\href
  {https://doi.org/10.1103/PhysRev.131.745} {\bibfield  {journal} {\bibinfo
  {journal} {Phys. Rev.}\ }\textbf {\bibinfo {volume} {131}},\ \bibinfo {pages}
  {745} (\bibinfo {year} {1963})}\BibitemShut {NoStop}%
\bibitem [{\citenamefont {Wiringa}\ \emph {et~al.}(1995)\citenamefont
  {Wiringa}, \citenamefont {Stoks},\ and\ \citenamefont
  {Schiavilla}}]{ref:Wiringa1995}%
  \BibitemOpen
  \bibfield  {author} {\bibinfo {author} {\bibfnamefont {R.~B.}\ \bibnamefont
  {Wiringa}}, \bibinfo {author} {\bibfnamefont {V.~G.~J.}\ \bibnamefont
  {Stoks}},\ and\ \bibinfo {author} {\bibfnamefont {R.}~\bibnamefont
  {Schiavilla}},\ }\bibfield  {title} {\bibinfo {title} {Accurate
  nucleon-nucleon potential with charge-independence breaking},\ }\href
  {https://doi.org/10.1103/PhysRevC.51.38} {\bibfield  {journal} {\bibinfo
  {journal} {Phys. Rev. C}\ }\textbf {\bibinfo {volume} {51}},\ \bibinfo
  {pages} {38} (\bibinfo {year} {1995})}\BibitemShut {NoStop}%
\bibitem [{\citenamefont {Ibbotson}\ \emph {et~al.}(1998)\citenamefont
  {Ibbotson}, \citenamefont {Glasmacher}, \citenamefont {Brown}, \citenamefont
  {Chen}, \citenamefont {DeCola}, \citenamefont {Groh}, \citenamefont {Honma},
  \citenamefont {Jewell}, \citenamefont {Kemper}, \citenamefont {Mantica},
  \citenamefont {Miller}, \citenamefont {Mueller}, \citenamefont {Otsuka},
  \citenamefont {Riley},\ and\ \citenamefont {Weber}}]{ref:Ibbotson1998}%
  \BibitemOpen
  \bibfield  {author} {\bibinfo {author} {\bibfnamefont {R.~W.}\ \bibnamefont
  {Ibbotson}}, \bibinfo {author} {\bibfnamefont {T.}~\bibnamefont
  {Glasmacher}}, \bibinfo {author} {\bibfnamefont {B.~A.}\ \bibnamefont
  {Brown}}, \bibinfo {author} {\bibfnamefont {L.}~\bibnamefont {Chen}},
  \bibinfo {author} {\bibfnamefont {M.~J.}\ \bibnamefont {DeCola}}, \bibinfo
  {author} {\bibfnamefont {D.~E.}\ \bibnamefont {Groh}}, \bibinfo {author}
  {\bibfnamefont {M.}~\bibnamefont {Honma}}, \bibinfo {author} {\bibfnamefont
  {C.}~\bibnamefont {Jewell}}, \bibinfo {author} {\bibfnamefont {K.~W.}\
  \bibnamefont {Kemper}}, \bibinfo {author} {\bibfnamefont {P.~F.}\
  \bibnamefont {Mantica}}, \bibinfo {author} {\bibfnamefont {D.}~\bibnamefont
  {Miller}}, \bibinfo {author} {\bibfnamefont {W.~F.}\ \bibnamefont {Mueller}},
  \bibinfo {author} {\bibfnamefont {T.}~\bibnamefont {Otsuka}}, \bibinfo
  {author} {\bibfnamefont {L.~A.}\ \bibnamefont {Riley}},\ and\ \bibinfo
  {author} {\bibfnamefont {M.}~\bibnamefont {Weber}},\ }\bibfield  {title}
  {\bibinfo {title} {Quadrupole collectivity in $^{32}\mathrm{Mg}$},\ }\href
  {https://doi.org/10.1103/PhysRevLett.80.2081} {\bibfield  {journal} {\bibinfo
   {journal} {Phys. Rev. Lett.}\ }\textbf {\bibinfo {volume} {80}},\ \bibinfo
  {pages} {2081} (\bibinfo {year} {1998})}\BibitemShut {NoStop}%
\bibitem [{\citenamefont {Glaudemans}\ \emph {et~al.}(1971)\citenamefont
  {Glaudemans}, \citenamefont {Endt},\ and\ \citenamefont
  {Dieperink}}]{ref:Glaudemans1971}%
  \BibitemOpen
  \bibfield  {author} {\bibinfo {author} {\bibfnamefont {P.~W.~M.}\
  \bibnamefont {Glaudemans}}, \bibinfo {author} {\bibfnamefont {P.~M.}\
  \bibnamefont {Endt}},\ and\ \bibinfo {author} {\bibfnamefont {A.~E.~L.}\
  \bibnamefont {Dieperink}},\ }\bibfield  {title} {\bibinfo {title}
  {Many-particle shell model calculation of electromagnetic transition rates
  and multipole moments in ${A} = 30$--$34$ nuclei},\ }\href
  {https://doi.org/10.1016/0003-4916(71)90299-5} {\bibfield  {journal}
  {\bibinfo  {journal} {Ann. Phys. (N.Y.)}\ }\textbf {\bibinfo {volume} {63}},\
  \bibinfo {pages} {134} (\bibinfo {year} {1971})}\BibitemShut {NoStop}%
\bibitem [{\citenamefont {Mackh}\ \emph {et~al.}(1973)\citenamefont {Mackh},
  \citenamefont {Oeschler}, \citenamefont {Wagner}, \citenamefont {Dehnhard},\
  and\ \citenamefont {Ohnuma}}]{ref:Mackh1973}%
  \BibitemOpen
  \bibfield  {author} {\bibinfo {author} {\bibfnamefont {H.}~\bibnamefont
  {Mackh}}, \bibinfo {author} {\bibfnamefont {H.}~\bibnamefont {Oeschler}},
  \bibinfo {author} {\bibfnamefont {G.~J.}\ \bibnamefont {Wagner}}, \bibinfo
  {author} {\bibfnamefont {D.}~\bibnamefont {Dehnhard}},\ and\ \bibinfo
  {author} {\bibfnamefont {H.}~\bibnamefont {Ohnuma}},\ }\bibfield  {title}
  {\bibinfo {title} {Energy levels in $^{30}${S}i from the
  $^{29}${S}i($\textit{d, p}$)$^{30}${S}i reaction},\ }\href
  {https://doi.org/10.1016/0375-9474(73)90412-1} {\bibfield  {journal}
  {\bibinfo  {journal} {Nucl. Phys. A}\ }\textbf {\bibinfo {volume} {202}},\
  \bibinfo {pages} {497} (\bibinfo {year} {1973})}\BibitemShut {NoStop}%
\bibitem [{\citenamefont {Baxter}\ \emph {et~al.}(1970)\citenamefont {Baxter},
  \citenamefont {Kuehner}, \citenamefont {Petty},\ and\ \citenamefont
  {O'Donnell}}]{ref:Baxter1970}%
  \BibitemOpen
  \bibfield  {author} {\bibinfo {author} {\bibfnamefont {A.~M.}\ \bibnamefont
  {Baxter}}, \bibinfo {author} {\bibfnamefont {J.~A.}\ \bibnamefont {Kuehner}},
  \bibinfo {author} {\bibfnamefont {D.~T.}\ \bibnamefont {Petty}},\ and\
  \bibinfo {author} {\bibfnamefont {J.~M.}\ \bibnamefont {O'Donnell}},\
  }\bibfield  {title} {\bibinfo {title} {Gamma-ray spectroscopy in
  $^{30}${S}i},\ }\href {https://doi.org/10.1103/PhysRevC.2.320} {\bibfield
  {journal} {\bibinfo  {journal} {Phys. Rev. C}\ }\textbf {\bibinfo {volume}
  {2}},\ \bibinfo {pages} {320} (\bibinfo {year} {1970})}\BibitemShut {NoStop}%
\bibitem [{\citenamefont {Bernstein}\ \emph {et~al.}(1979)\citenamefont
  {Bernstein}, \citenamefont {Brown},\ and\ \citenamefont
  {Madsen}}]{ref:Bernstein1979}%
  \BibitemOpen
  \bibfield  {author} {\bibinfo {author} {\bibfnamefont {A.~M.}\ \bibnamefont
  {Bernstein}}, \bibinfo {author} {\bibfnamefont {V.~R.}\ \bibnamefont
  {Brown}},\ and\ \bibinfo {author} {\bibfnamefont {V.~A.}\ \bibnamefont
  {Madsen}},\ }\bibfield  {title} {\bibinfo {title} {Isospin decomposition of
  nuclear multipole matrix elements from $\gamma$ decay rates of mirror
  transitions: Test of values obtained with hadronic probes},\ }\href
  {https://doi.org/10.1103/PhysRevLett.42.425} {\bibfield  {journal} {\bibinfo
  {journal} {Phys. Rev. Lett.}\ }\textbf {\bibinfo {volume} {42}},\ \bibinfo
  {pages} {425} (\bibinfo {year} {1979})}\BibitemShut {NoStop}%
\bibitem [{\citenamefont {Mutschler}\ \emph {et~al.}(2017)\citenamefont
  {Mutschler}, \citenamefont {Lemasson}, \citenamefont {Sorlin}, \citenamefont
  {Bazin}, \citenamefont {Borcea}, \citenamefont {Borcea}, \citenamefont
  {Dombr{\'a}di}, \citenamefont {Ebran}, \citenamefont {Gade}, \citenamefont
  {Iwasaki}, \citenamefont {Khan}, \citenamefont {Lepailleur}, \citenamefont
  {Recchia}, \citenamefont {Roger}, \citenamefont {Rotaru}, \citenamefont
  {Sohler}, \citenamefont {Stanoiu}, \citenamefont {Stroberg}, \citenamefont
  {Tostevin}, \citenamefont {Vandebrouck}, \citenamefont {Weisshaar},\ and\
  \citenamefont {Wimmer}}]{ref:Mutschler2017}%
  \BibitemOpen
  \bibfield  {author} {\bibinfo {author} {\bibfnamefont {A.}~\bibnamefont
  {Mutschler}}, \bibinfo {author} {\bibfnamefont {A.}~\bibnamefont {Lemasson}},
  \bibinfo {author} {\bibfnamefont {O.}~\bibnamefont {Sorlin}}, \bibinfo
  {author} {\bibfnamefont {D.}~\bibnamefont {Bazin}}, \bibinfo {author}
  {\bibfnamefont {C.}~\bibnamefont {Borcea}}, \bibinfo {author} {\bibfnamefont
  {R.}~\bibnamefont {Borcea}}, \bibinfo {author} {\bibfnamefont
  {Z.}~\bibnamefont {Dombr{\'a}di}}, \bibinfo {author} {\bibfnamefont {J.~P.}\
  \bibnamefont {Ebran}}, \bibinfo {author} {\bibfnamefont {A.}~\bibnamefont
  {Gade}}, \bibinfo {author} {\bibfnamefont {H.}~\bibnamefont {Iwasaki}},
  \bibinfo {author} {\bibfnamefont {E.}~\bibnamefont {Khan}}, \bibinfo {author}
  {\bibfnamefont {A.}~\bibnamefont {Lepailleur}}, \bibinfo {author}
  {\bibfnamefont {F.}~\bibnamefont {Recchia}}, \bibinfo {author} {\bibfnamefont
  {T.}~\bibnamefont {Roger}}, \bibinfo {author} {\bibfnamefont
  {F.}~\bibnamefont {Rotaru}}, \bibinfo {author} {\bibfnamefont
  {D.}~\bibnamefont {Sohler}}, \bibinfo {author} {\bibfnamefont
  {M.}~\bibnamefont {Stanoiu}}, \bibinfo {author} {\bibfnamefont {S.~R.}\
  \bibnamefont {Stroberg}}, \bibinfo {author} {\bibfnamefont {J.~A.}\
  \bibnamefont {Tostevin}}, \bibinfo {author} {\bibfnamefont {M.}~\bibnamefont
  {Vandebrouck}}, \bibinfo {author} {\bibfnamefont {D.}~\bibnamefont
  {Weisshaar}},\ and\ \bibinfo {author} {\bibfnamefont {K.}~\bibnamefont
  {Wimmer}},\ }\bibfield  {title} {\bibinfo {title} {A proton density bubble in
  the doubly magic $^{34}\mathrm{Si}$ nucleus},\ }\href
  {https://doi.org/10.1038/nphys3916} {\bibfield  {journal} {\bibinfo
  {journal} {Nature Physics}\ }\textbf {\bibinfo {volume} {13}},\ \bibinfo
  {pages} {152} (\bibinfo {year} {2017})}\BibitemShut {NoStop}%
\bibitem [{\citenamefont {Baumann}\ \emph {et~al.}(1989)\citenamefont
  {Baumann}, \citenamefont {Huck}, \citenamefont {Klotz}, \citenamefont
  {Knipper}, \citenamefont {Walter}, \citenamefont {Marguier}, \citenamefont
  {Ravn}, \citenamefont {Richard-Serre}, \citenamefont {Poves},\ and\
  \citenamefont {Retamosa}}]{ref:Baumann1989}%
  \BibitemOpen
  \bibfield  {author} {\bibinfo {author} {\bibfnamefont {P.}~\bibnamefont
  {Baumann}}, \bibinfo {author} {\bibfnamefont {A.}~\bibnamefont {Huck}},
  \bibinfo {author} {\bibfnamefont {G.}~\bibnamefont {Klotz}}, \bibinfo
  {author} {\bibfnamefont {A.}~\bibnamefont {Knipper}}, \bibinfo {author}
  {\bibfnamefont {G.}~\bibnamefont {Walter}}, \bibinfo {author} {\bibfnamefont
  {G.}~\bibnamefont {Marguier}}, \bibinfo {author} {\bibfnamefont {H.~L.}\
  \bibnamefont {Ravn}}, \bibinfo {author} {\bibfnamefont {C.}~\bibnamefont
  {Richard-Serre}}, \bibinfo {author} {\bibfnamefont {A.}~\bibnamefont
  {Poves}},\ and\ \bibinfo {author} {\bibfnamefont {J.}~\bibnamefont
  {Retamosa}},\ }\bibfield  {title} {\bibinfo {title} {$^{34}\mathrm{Si}$:
  $\mathrm{A}$ new doubly magic nucleus?},\ }\href
  {https://doi.org/https://doi.org/10.1016/0370-2693(89)90974-X} {\bibfield
  {journal} {\bibinfo  {journal} {Phys. Lett. B}\ }\textbf {\bibinfo {volume}
  {228}},\ \bibinfo {pages} {458} (\bibinfo {year} {1989})}\BibitemShut
  {NoStop}%
\bibitem [{\citenamefont {Nummela}\ \emph {et~al.}(2001)\citenamefont
  {Nummela}, \citenamefont {Baumann}, \citenamefont {Caurier}, \citenamefont
  {Dessagne}, \citenamefont {Jokinen}, \citenamefont {Knipper}, \citenamefont
  {Le~Scornet}, \citenamefont {Mieh\'e}, \citenamefont {Nowacki}, \citenamefont
  {Oinonen}, \citenamefont {Radivojevic}, \citenamefont {Ramdhane},
  \citenamefont {Walter},\ and\ \citenamefont {\"Ayst\"o}}]{ref:Nummela2001}%
  \BibitemOpen
  \bibfield  {author} {\bibinfo {author} {\bibfnamefont {S.}~\bibnamefont
  {Nummela}}, \bibinfo {author} {\bibfnamefont {P.}~\bibnamefont {Baumann}},
  \bibinfo {author} {\bibfnamefont {E.}~\bibnamefont {Caurier}}, \bibinfo
  {author} {\bibfnamefont {P.}~\bibnamefont {Dessagne}}, \bibinfo {author}
  {\bibfnamefont {A.}~\bibnamefont {Jokinen}}, \bibinfo {author} {\bibfnamefont
  {A.}~\bibnamefont {Knipper}}, \bibinfo {author} {\bibfnamefont
  {G.}~\bibnamefont {Le~Scornet}}, \bibinfo {author} {\bibfnamefont
  {C.}~\bibnamefont {Mieh\'e}}, \bibinfo {author} {\bibfnamefont
  {F.}~\bibnamefont {Nowacki}}, \bibinfo {author} {\bibfnamefont
  {M.}~\bibnamefont {Oinonen}}, \bibinfo {author} {\bibfnamefont
  {Z.}~\bibnamefont {Radivojevic}}, \bibinfo {author} {\bibfnamefont
  {M.}~\bibnamefont {Ramdhane}}, \bibinfo {author} {\bibfnamefont
  {G.}~\bibnamefont {Walter}},\ and\ \bibinfo {author} {\bibfnamefont
  {J.}~\bibnamefont {\"Ayst\"o}},\ }\bibfield  {title} {\bibinfo {title}
  {Spectroscopy of ${}^{34,35}\mathrm{Si}$ by $\ensuremath{\beta}$ decay:
  $sd\ensuremath{-}fp$ shell gap and single-particle states},\ }\href
  {https://doi.org/10.1103/PhysRevC.63.044316} {\bibfield  {journal} {\bibinfo
  {journal} {Phys. Rev. C}\ }\textbf {\bibinfo {volume} {63}},\ \bibinfo
  {pages} {044316} (\bibinfo {year} {2001})}\BibitemShut {NoStop}%
\bibitem [{\citenamefont {Paschalis}\ \emph {et~al.}(2011)\citenamefont
  {Paschalis}, \citenamefont {Fallon}, \citenamefont {Macchiavelli},
  \citenamefont {Clark}, \citenamefont {Crawford}, \citenamefont {Lister},
  \citenamefont {Vondrasek}, \citenamefont {Bazin}, \citenamefont {Berryman},
  \citenamefont {Bertulani}, \citenamefont {Brown}, \citenamefont {Campbell},
  \citenamefont {Carpenter}, \citenamefont {Chapman}, \citenamefont
  {Chowdhury}, \citenamefont {Gade}, \citenamefont {Glasmacher}, \citenamefont
  {Greene}, \citenamefont {Hecht}, \citenamefont {Hoffman}, \citenamefont
  {Janssens}, \citenamefont {Kay}, \citenamefont {Lauritsen}, \citenamefont
  {Nair}, \citenamefont {Ong}, \citenamefont {Reviol}, \citenamefont
  {Schiffer}, \citenamefont {Seweryniak}, \citenamefont {Varner}, \citenamefont
  {Weisshaar}, \citenamefont {Wiedenh{\"o}ver},\ and\ \citenamefont
  {Zhu}}]{ref:Paschalis2011}%
  \BibitemOpen
  \bibfield  {author} {\bibinfo {author} {\bibfnamefont {S.}~\bibnamefont
  {Paschalis}}, \bibinfo {author} {\bibfnamefont {P.}~\bibnamefont {Fallon}},
  \bibinfo {author} {\bibfnamefont {A.~O.}\ \bibnamefont {Macchiavelli}},
  \bibinfo {author} {\bibfnamefont {R.~M.}\ \bibnamefont {Clark}}, \bibinfo
  {author} {\bibfnamefont {H.~L.}\ \bibnamefont {Crawford}}, \bibinfo {author}
  {\bibfnamefont {C.~J.}\ \bibnamefont {Lister}}, \bibinfo {author}
  {\bibfnamefont {R.}~\bibnamefont {Vondrasek}}, \bibinfo {author}
  {\bibfnamefont {D.}~\bibnamefont {Bazin}}, \bibinfo {author} {\bibfnamefont
  {J.~S.}\ \bibnamefont {Berryman}}, \bibinfo {author} {\bibfnamefont {C.~A.}\
  \bibnamefont {Bertulani}}, \bibinfo {author} {\bibfnamefont {B.~A.}\
  \bibnamefont {Brown}}, \bibinfo {author} {\bibfnamefont {C.~M.}\ \bibnamefont
  {Campbell}}, \bibinfo {author} {\bibfnamefont {M.~P.}\ \bibnamefont
  {Carpenter}}, \bibinfo {author} {\bibfnamefont {R.}~\bibnamefont {Chapman}},
  \bibinfo {author} {\bibfnamefont {P.}~\bibnamefont {Chowdhury}}, \bibinfo
  {author} {\bibfnamefont {A.}~\bibnamefont {Gade}}, \bibinfo {author}
  {\bibfnamefont {T.}~\bibnamefont {Glasmacher}}, \bibinfo {author}
  {\bibfnamefont {J.~P.}\ \bibnamefont {Greene}}, \bibinfo {author}
  {\bibfnamefont {A.~A.}\ \bibnamefont {Hecht}}, \bibinfo {author}
  {\bibfnamefont {C.~R.}\ \bibnamefont {Hoffman}}, \bibinfo {author}
  {\bibfnamefont {R.~V.~F.}\ \bibnamefont {Janssens}}, \bibinfo {author}
  {\bibfnamefont {B.~P.}\ \bibnamefont {Kay}}, \bibinfo {author} {\bibfnamefont
  {T.}~\bibnamefont {Lauritsen}}, \bibinfo {author} {\bibfnamefont
  {C.}~\bibnamefont {Nair}}, \bibinfo {author} {\bibfnamefont {H.~J.}\
  \bibnamefont {Ong}}, \bibinfo {author} {\bibfnamefont {W.}~\bibnamefont
  {Reviol}}, \bibinfo {author} {\bibfnamefont {J.~P.}\ \bibnamefont
  {Schiffer}}, \bibinfo {author} {\bibfnamefont {D.}~\bibnamefont
  {Seweryniak}}, \bibinfo {author} {\bibfnamefont {R.~L.}\ \bibnamefont
  {Varner}}, \bibinfo {author} {\bibfnamefont {D.}~\bibnamefont {Weisshaar}},
  \bibinfo {author} {\bibfnamefont {I.}~\bibnamefont {Wiedenh{\"o}ver}},\ and\
  \bibinfo {author} {\bibfnamefont {S.}~\bibnamefont {Zhu}},\ }\bibfield
  {title} {\bibinfo {title} {The deformed 0$^+$ state in $^{34}\mathrm{Si}$},\
  }\href {https://doi.org/10.1088/1742-6596/312/9/092050} {\bibfield  {journal}
  {\bibinfo  {journal} {Journal of Physics: Conference Series}\ }\textbf
  {\bibinfo {volume} {312}},\ \bibinfo {pages} {092050} (\bibinfo {year}
  {2011})}\BibitemShut {NoStop}%
\bibitem [{\citenamefont {Rotaru}\ \emph {et~al.}(2012)\citenamefont {Rotaru},
  \citenamefont {Negoita}, \citenamefont {Gr\'evy}, \citenamefont {Mrazek},
  \citenamefont {Lukyanov}, \citenamefont {Nowacki}, \citenamefont {Poves},
  \citenamefont {Sorlin}, \citenamefont {Borcea}, \citenamefont {Borcea},
  \citenamefont {Buta}, \citenamefont {C\'aceres}, \citenamefont {Calinescu},
  \citenamefont {Chevrier}, \citenamefont {Dombr\'adi}, \citenamefont {Daugas},
  \citenamefont {Lebhertz}, \citenamefont {Penionzhkevich}, \citenamefont
  {Petrone}, \citenamefont {Sohler}, \citenamefont {Stanoiu},\ and\
  \citenamefont {Thomas}}]{ref:Rotaru2012}%
  \BibitemOpen
  \bibfield  {author} {\bibinfo {author} {\bibfnamefont {F.}~\bibnamefont
  {Rotaru}}, \bibinfo {author} {\bibfnamefont {F.}~\bibnamefont {Negoita}},
  \bibinfo {author} {\bibfnamefont {S.}~\bibnamefont {Gr\'evy}}, \bibinfo
  {author} {\bibfnamefont {J.}~\bibnamefont {Mrazek}}, \bibinfo {author}
  {\bibfnamefont {S.}~\bibnamefont {Lukyanov}}, \bibinfo {author}
  {\bibfnamefont {F.}~\bibnamefont {Nowacki}}, \bibinfo {author} {\bibfnamefont
  {A.}~\bibnamefont {Poves}}, \bibinfo {author} {\bibfnamefont
  {O.}~\bibnamefont {Sorlin}}, \bibinfo {author} {\bibfnamefont
  {C.}~\bibnamefont {Borcea}}, \bibinfo {author} {\bibfnamefont
  {R.}~\bibnamefont {Borcea}}, \bibinfo {author} {\bibfnamefont
  {A.}~\bibnamefont {Buta}}, \bibinfo {author} {\bibfnamefont {L.}~\bibnamefont
  {C\'aceres}}, \bibinfo {author} {\bibfnamefont {S.}~\bibnamefont
  {Calinescu}}, \bibinfo {author} {\bibfnamefont {R.}~\bibnamefont {Chevrier}},
  \bibinfo {author} {\bibfnamefont {Z.}~\bibnamefont {Dombr\'adi}}, \bibinfo
  {author} {\bibfnamefont {J.~M.}\ \bibnamefont {Daugas}}, \bibinfo {author}
  {\bibfnamefont {D.}~\bibnamefont {Lebhertz}}, \bibinfo {author}
  {\bibfnamefont {Y.}~\bibnamefont {Penionzhkevich}}, \bibinfo {author}
  {\bibfnamefont {C.}~\bibnamefont {Petrone}}, \bibinfo {author} {\bibfnamefont
  {D.}~\bibnamefont {Sohler}}, \bibinfo {author} {\bibfnamefont
  {M.}~\bibnamefont {Stanoiu}},\ and\ \bibinfo {author} {\bibfnamefont {J.~C.}\
  \bibnamefont {Thomas}},\ }\bibfield  {title} {\bibinfo {title} {Unveiling the
  intruder deformed ${0}_{2}^{+}$ state in $^{34}\mathrm{Si}$},\ }\href
  {https://doi.org/10.1103/PhysRevLett.109.092503} {\bibfield  {journal}
  {\bibinfo  {journal} {Phys. Rev. Lett.}\ }\textbf {\bibinfo {volume} {109}},\
  \bibinfo {pages} {092503} (\bibinfo {year} {2012})}\BibitemShut {NoStop}%
\bibitem [{\citenamefont {Lic\ifmmode~\u{a}\else \u{a}\fi{}}\ \emph
  {et~al.}(2019)\citenamefont {Lic\ifmmode~\u{a}\else \u{a}\fi{}},
  \citenamefont {Rotaru}, \citenamefont {Borge}, \citenamefont {Gr\'evy},
  \citenamefont {Negoi\ifmmode \mbox{\c{t}}\else \c{t}\fi{}\ifmmode~\u{a}\else
  \u{a}\fi{}}, \citenamefont {Poves}, \citenamefont {Sorlin}, \citenamefont
  {Andreyev}, \citenamefont {Borcea}, \citenamefont {Costache}, \citenamefont
  {De~Witte}, \citenamefont {Fraile}, \citenamefont {Greenlees}, \citenamefont
  {Huyse}, \citenamefont {Ionescu}, \citenamefont {Kisyov}, \citenamefont
  {Konki}, \citenamefont {Lazarus}, \citenamefont {Madurga}, \citenamefont
  {M\ifmmode~\u{a}\else \u{a}\fi{}rginean}, \citenamefont {M\ifmmode~\u{a}\else
  \u{a}\fi{}rginean}, \citenamefont {Mihai}, \citenamefont {Mihai},
  \citenamefont {Negret}, \citenamefont {Nowacki}, \citenamefont {Page},
  \citenamefont {Pakarinen}, \citenamefont {Pucknell}, \citenamefont {Rahkila},
  \citenamefont {Rapisarda}, \citenamefont {\ifmmode~\mbox{\c{S}}\else
  \c{S}\fi{}erban}, \citenamefont {Sotty}, \citenamefont {Stan}, \citenamefont
  {St\ifmmode~\u{a}\else \u{a}\fi{}noiu}, \citenamefont {Tengblad},
  \citenamefont {Turturic\ifmmode~\u{a}\else \u{a}\fi{}}, \citenamefont
  {Van~Duppen}, \citenamefont {Warr}, \citenamefont {Dessagne}, \citenamefont
  {Stora}, \citenamefont {Borcea}, \citenamefont {C\ifmmode~\u{a}\else
  \u{a}\fi{}linescu}, \citenamefont {Daugas}, \citenamefont {Filipescu},
  \citenamefont {Kuti}, \citenamefont {Franchoo}, \citenamefont {Gheorghe},
  \citenamefont {Morfouace}, \citenamefont {Morel}, \citenamefont {Mrazek},
  \citenamefont {Pietreanu}, \citenamefont {Sohler}, \citenamefont {Stefan},
  \citenamefont {\ifmmode \mbox{\c{S}}\else \c{S}\fi{}uv\ifmmode \u{a}\else
  \u{a}\fi{}il\ifmmode~\u{a}\else \u{a}\fi{}}, \citenamefont {Toma},\ and\
  \citenamefont {Ur}}]{ref:Lica2019}%
  \BibitemOpen
  \bibfield  {author} {\bibinfo {author} {\bibfnamefont {R.}~\bibnamefont
  {Lic\ifmmode~\u{a}\else \u{a}\fi{}}}, \bibinfo {author} {\bibfnamefont
  {F.}~\bibnamefont {Rotaru}}, \bibinfo {author} {\bibfnamefont {M.~J.~G.}\
  \bibnamefont {Borge}}, \bibinfo {author} {\bibfnamefont {S.}~\bibnamefont
  {Gr\'evy}}, \bibinfo {author} {\bibfnamefont {F.}~\bibnamefont {Negoi\ifmmode
  \mbox{\c{t}}\else \c{t}\fi{}\ifmmode~\u{a}\else \u{a}\fi{}}}, \bibinfo
  {author} {\bibfnamefont {A.}~\bibnamefont {Poves}}, \bibinfo {author}
  {\bibfnamefont {O.}~\bibnamefont {Sorlin}}, \bibinfo {author} {\bibfnamefont
  {A.~N.}\ \bibnamefont {Andreyev}}, \bibinfo {author} {\bibfnamefont
  {R.}~\bibnamefont {Borcea}}, \bibinfo {author} {\bibfnamefont
  {C.}~\bibnamefont {Costache}}, \bibinfo {author} {\bibfnamefont
  {H.}~\bibnamefont {De~Witte}}, \bibinfo {author} {\bibfnamefont {L.~M.}\
  \bibnamefont {Fraile}}, \bibinfo {author} {\bibfnamefont {P.~T.}\
  \bibnamefont {Greenlees}}, \bibinfo {author} {\bibfnamefont {M.}~\bibnamefont
  {Huyse}}, \bibinfo {author} {\bibfnamefont {A.}~\bibnamefont {Ionescu}},
  \bibinfo {author} {\bibfnamefont {S.}~\bibnamefont {Kisyov}}, \bibinfo
  {author} {\bibfnamefont {J.}~\bibnamefont {Konki}}, \bibinfo {author}
  {\bibfnamefont {I.}~\bibnamefont {Lazarus}}, \bibinfo {author} {\bibfnamefont
  {M.}~\bibnamefont {Madurga}}, \bibinfo {author} {\bibfnamefont
  {N.}~\bibnamefont {M\ifmmode~\u{a}\else \u{a}\fi{}rginean}}, \bibinfo
  {author} {\bibfnamefont {R.}~\bibnamefont {M\ifmmode~\u{a}\else
  \u{a}\fi{}rginean}}, \bibinfo {author} {\bibfnamefont {C.}~\bibnamefont
  {Mihai}}, \bibinfo {author} {\bibfnamefont {R.~E.}\ \bibnamefont {Mihai}},
  \bibinfo {author} {\bibfnamefont {A.}~\bibnamefont {Negret}}, \bibinfo
  {author} {\bibfnamefont {F.}~\bibnamefont {Nowacki}}, \bibinfo {author}
  {\bibfnamefont {R.~D.}\ \bibnamefont {Page}}, \bibinfo {author}
  {\bibfnamefont {J.}~\bibnamefont {Pakarinen}}, \bibinfo {author}
  {\bibfnamefont {V.}~\bibnamefont {Pucknell}}, \bibinfo {author}
  {\bibfnamefont {P.}~\bibnamefont {Rahkila}}, \bibinfo {author} {\bibfnamefont
  {E.}~\bibnamefont {Rapisarda}}, \bibinfo {author} {\bibfnamefont
  {A.}~\bibnamefont {\ifmmode~\mbox{\c{S}}\else \c{S}\fi{}erban}}, \bibinfo
  {author} {\bibfnamefont {C.~O.}\ \bibnamefont {Sotty}}, \bibinfo {author}
  {\bibfnamefont {L.}~\bibnamefont {Stan}}, \bibinfo {author} {\bibfnamefont
  {M.}~\bibnamefont {St\ifmmode~\u{a}\else \u{a}\fi{}noiu}}, \bibinfo {author}
  {\bibfnamefont {O.}~\bibnamefont {Tengblad}}, \bibinfo {author}
  {\bibfnamefont {A.}~\bibnamefont {Turturic\ifmmode~\u{a}\else \u{a}\fi{}}},
  \bibinfo {author} {\bibfnamefont {P.}~\bibnamefont {Van~Duppen}}, \bibinfo
  {author} {\bibfnamefont {N.}~\bibnamefont {Warr}}, \bibinfo {author}
  {\bibfnamefont {P.}~\bibnamefont {Dessagne}}, \bibinfo {author}
  {\bibfnamefont {T.}~\bibnamefont {Stora}}, \bibinfo {author} {\bibfnamefont
  {C.}~\bibnamefont {Borcea}}, \bibinfo {author} {\bibfnamefont
  {S.}~\bibnamefont {C\ifmmode~\u{a}\else \u{a}\fi{}linescu}}, \bibinfo
  {author} {\bibfnamefont {J.~M.}\ \bibnamefont {Daugas}}, \bibinfo {author}
  {\bibfnamefont {D.}~\bibnamefont {Filipescu}}, \bibinfo {author}
  {\bibfnamefont {I.}~\bibnamefont {Kuti}}, \bibinfo {author} {\bibfnamefont
  {S.}~\bibnamefont {Franchoo}}, \bibinfo {author} {\bibfnamefont
  {I.}~\bibnamefont {Gheorghe}}, \bibinfo {author} {\bibfnamefont
  {P.}~\bibnamefont {Morfouace}}, \bibinfo {author} {\bibfnamefont
  {P.}~\bibnamefont {Morel}}, \bibinfo {author} {\bibfnamefont
  {J.}~\bibnamefont {Mrazek}}, \bibinfo {author} {\bibfnamefont
  {D.}~\bibnamefont {Pietreanu}}, \bibinfo {author} {\bibfnamefont
  {D.}~\bibnamefont {Sohler}}, \bibinfo {author} {\bibfnamefont
  {I.}~\bibnamefont {Stefan}}, \bibinfo {author} {\bibfnamefont
  {R.}~\bibnamefont {\ifmmode \mbox{\c{S}}\else \c{S}\fi{}uv\ifmmode \u{a}\else
  \u{a}\fi{}il\ifmmode~\u{a}\else \u{a}\fi{}}}, \bibinfo {author}
  {\bibfnamefont {S.}~\bibnamefont {Toma}},\ and\ \bibinfo {author}
  {\bibfnamefont {C.~A.}\ \bibnamefont {Ur}} (\bibinfo {collaboration} {IDS
  Collaboration}),\ }\bibfield  {title} {\bibinfo {title} {Normal and intruder
  configurations in $^{34}\mathrm{Si}$ populated in the
  ${\ensuremath{\beta}}^{\ensuremath{-}}$ decay of $^{34}\mathrm{Mg}$ and
  $^{34}\mathrm{Al}$},\ }\href {https://doi.org/10.1103/PhysRevC.100.034306}
  {\bibfield  {journal} {\bibinfo  {journal} {Phys. Rev. C}\ }\textbf {\bibinfo
  {volume} {100}},\ \bibinfo {pages} {034306} (\bibinfo {year}
  {2019})}\BibitemShut {NoStop}%
\bibitem [{\citenamefont {Grocutt}\ \emph {et~al.}(2022)\citenamefont
  {Grocutt}, \citenamefont {Chapman}, \citenamefont {Bouhelal}, \citenamefont
  {Haas}, \citenamefont {Goasduff}, \citenamefont {Smith}, \citenamefont
  {Lubna}, \citenamefont {Courtin}, \citenamefont {Bazzacco}, \citenamefont
  {Braunroth}, \citenamefont {Capponi}, \citenamefont {Corradi}, \citenamefont
  {Derkx}, \citenamefont {Desesquelles}, \citenamefont {Doncel}, \citenamefont
  {Fioretto}, \citenamefont {Gottardo}, \citenamefont {Liberati}, \citenamefont
  {Melon}, \citenamefont {Mengoni}, \citenamefont {Michelagnoli}, \citenamefont
  {Mijatovi\ifmmode~\acute{c}\else \'{c}\fi{}}, \citenamefont {Modamio},
  \citenamefont {Montagnoli}, \citenamefont {Montanari}, \citenamefont
  {Mulholland}, \citenamefont {Napoli}, \citenamefont {Petrache}, \citenamefont
  {Pipidis}, \citenamefont {Recchia}, \citenamefont {Sahin}, \citenamefont
  {Singh}, \citenamefont {Stefanini}, \citenamefont {Szilner},\ and\
  \citenamefont {Valiente-Dob\'on}}]{ref:Grocutt2022}%
  \BibitemOpen
  \bibfield  {author} {\bibinfo {author} {\bibfnamefont {L.}~\bibnamefont
  {Grocutt}}, \bibinfo {author} {\bibfnamefont {R.}~\bibnamefont {Chapman}},
  \bibinfo {author} {\bibfnamefont {M.}~\bibnamefont {Bouhelal}}, \bibinfo
  {author} {\bibfnamefont {F.}~\bibnamefont {Haas}}, \bibinfo {author}
  {\bibfnamefont {A.}~\bibnamefont {Goasduff}}, \bibinfo {author}
  {\bibfnamefont {J.~F.}\ \bibnamefont {Smith}}, \bibinfo {author}
  {\bibfnamefont {R.~S.}\ \bibnamefont {Lubna}}, \bibinfo {author}
  {\bibfnamefont {S.}~\bibnamefont {Courtin}}, \bibinfo {author} {\bibfnamefont
  {D.}~\bibnamefont {Bazzacco}}, \bibinfo {author} {\bibfnamefont
  {T.}~\bibnamefont {Braunroth}}, \bibinfo {author} {\bibfnamefont
  {L.}~\bibnamefont {Capponi}}, \bibinfo {author} {\bibfnamefont
  {L.}~\bibnamefont {Corradi}}, \bibinfo {author} {\bibfnamefont
  {X.}~\bibnamefont {Derkx}}, \bibinfo {author} {\bibfnamefont
  {P.}~\bibnamefont {Desesquelles}}, \bibinfo {author} {\bibfnamefont
  {M.}~\bibnamefont {Doncel}}, \bibinfo {author} {\bibfnamefont
  {E.}~\bibnamefont {Fioretto}}, \bibinfo {author} {\bibfnamefont
  {A.}~\bibnamefont {Gottardo}}, \bibinfo {author} {\bibfnamefont
  {V.}~\bibnamefont {Liberati}}, \bibinfo {author} {\bibfnamefont
  {B.}~\bibnamefont {Melon}}, \bibinfo {author} {\bibfnamefont
  {D.}~\bibnamefont {Mengoni}}, \bibinfo {author} {\bibfnamefont
  {C.}~\bibnamefont {Michelagnoli}}, \bibinfo {author} {\bibfnamefont
  {T.}~\bibnamefont {Mijatovi\ifmmode~\acute{c}\else \'{c}\fi{}}}, \bibinfo
  {author} {\bibfnamefont {V.}~\bibnamefont {Modamio}}, \bibinfo {author}
  {\bibfnamefont {G.}~\bibnamefont {Montagnoli}}, \bibinfo {author}
  {\bibfnamefont {D.}~\bibnamefont {Montanari}}, \bibinfo {author}
  {\bibfnamefont {K.~F.}\ \bibnamefont {Mulholland}}, \bibinfo {author}
  {\bibfnamefont {D.~R.}\ \bibnamefont {Napoli}}, \bibinfo {author}
  {\bibfnamefont {C.~M.}\ \bibnamefont {Petrache}}, \bibinfo {author}
  {\bibfnamefont {A.}~\bibnamefont {Pipidis}}, \bibinfo {author} {\bibfnamefont
  {F.}~\bibnamefont {Recchia}}, \bibinfo {author} {\bibfnamefont
  {E.}~\bibnamefont {Sahin}}, \bibinfo {author} {\bibfnamefont {P.~P.}\
  \bibnamefont {Singh}}, \bibinfo {author} {\bibfnamefont {A.~M.}\ \bibnamefont
  {Stefanini}}, \bibinfo {author} {\bibfnamefont {S.}~\bibnamefont {Szilner}},\
  and\ \bibinfo {author} {\bibfnamefont {J.~J.}\ \bibnamefont
  {Valiente-Dob\'on}},\ }\bibfield  {title} {\bibinfo {title} {Lifetime
  measurements of states of $^{35}\mathrm{S}$, $^{36}\mathrm{S}$,
  $^{37}\mathrm{S}$, and $^{38}\mathrm{S}$ using the $\textrm{AGATA}$
  $\ensuremath{\gamma}$-ray tracking spectrometer},\ }\href
  {https://doi.org/10.1103/PhysRevC.106.024314} {\bibfield  {journal} {\bibinfo
   {journal} {Phys. Rev. C}\ }\textbf {\bibinfo {volume} {106}},\ \bibinfo
  {pages} {024314} (\bibinfo {year} {2022})}\BibitemShut {NoStop}%
\bibitem [{\citenamefont {Rudolph}\ \emph {et~al.}(2002)\citenamefont
  {Rudolph}, \citenamefont {Baktash}, \citenamefont {Dobaczewski},
  \citenamefont {Nazarewicz}, \citenamefont {Satula}, \citenamefont {Devlin},
  \citenamefont {Hartley}, \citenamefont {LaFosse}, \citenamefont {Lerma},
  \citenamefont {Sarantites}, \citenamefont {Afanasjev}, \citenamefont
  {Carpenter}, \citenamefont {Lauritsen}, \citenamefont {Macchiavelli},
  \citenamefont {Svensson}, \citenamefont {Andreoiu}, \citenamefont {Axiotis},
  \citenamefont {Farnea}, \citenamefont {Gadea}, \citenamefont {Napoli},\ and\
  \citenamefont {de~Angelis}}]{ref:Rudolph2002}%
  \BibitemOpen
  \bibfield  {author} {\bibinfo {author} {\bibfnamefont {D.}~\bibnamefont
  {Rudolph}}, \bibinfo {author} {\bibfnamefont {C.}~\bibnamefont {Baktash}},
  \bibinfo {author} {\bibfnamefont {J.}~\bibnamefont {Dobaczewski}}, \bibinfo
  {author} {\bibfnamefont {W.}~\bibnamefont {Nazarewicz}}, \bibinfo {author}
  {\bibfnamefont {W.}~\bibnamefont {Satula}}, \bibinfo {author} {\bibfnamefont
  {M.}~\bibnamefont {Devlin}}, \bibinfo {author} {\bibfnamefont {D.~R.}\
  \bibnamefont {Hartley}}, \bibinfo {author} {\bibfnamefont {D.~R.}\
  \bibnamefont {LaFosse}}, \bibinfo {author} {\bibfnamefont {F.}~\bibnamefont
  {Lerma}}, \bibinfo {author} {\bibfnamefont {D.~G.}\ \bibnamefont
  {Sarantites}}, \bibinfo {author} {\bibfnamefont {A.~V.}\ \bibnamefont
  {Afanasjev}}, \bibinfo {author} {\bibfnamefont {M.~P.}\ \bibnamefont
  {Carpenter}}, \bibinfo {author} {\bibfnamefont {T.}~\bibnamefont
  {Lauritsen}}, \bibinfo {author} {\bibfnamefont {A.~O.}\ \bibnamefont
  {Macchiavelli}}, \bibinfo {author} {\bibfnamefont {C.~E.}\ \bibnamefont
  {Svensson}}, \bibinfo {author} {\bibfnamefont {C.}~\bibnamefont {Andreoiu}},
  \bibinfo {author} {\bibfnamefont {M.}~\bibnamefont {Axiotis}}, \bibinfo
  {author} {\bibfnamefont {E.}~\bibnamefont {Farnea}}, \bibinfo {author}
  {\bibfnamefont {A.}~\bibnamefont {Gadea}}, \bibinfo {author} {\bibfnamefont
  {D.~R.}\ \bibnamefont {Napoli}},\ and\ \bibinfo {author} {\bibfnamefont
  {G.}~\bibnamefont {de~Angelis}},\ }\bibfield  {title} {\bibinfo {title}
  {Spherical and deformed high-spin states in $^{38}\mathrm{Ar}$},\ }\href
  {https://doi.org/10.1103/PhysRevC.65.034305} {\bibfield  {journal} {\bibinfo
  {journal} {Phys. Rev. C}\ }\textbf {\bibinfo {volume} {65}},\ \bibinfo
  {pages} {034305} (\bibinfo {year} {2002})}\BibitemShut {NoStop}%
\bibitem [{\citenamefont {Engelbertink}\ \emph {et~al.}(1970)\citenamefont
  {Engelbertink}, \citenamefont {Jones}, \citenamefont {Olness},\ and\
  \citenamefont {Warburton}}]{ref:Engelbertink1970}%
  \BibitemOpen
  \bibfield  {author} {\bibinfo {author} {\bibfnamefont {G.~A.~P.}\
  \bibnamefont {Engelbertink}}, \bibinfo {author} {\bibfnamefont {K.~W.}\
  \bibnamefont {Jones}}, \bibinfo {author} {\bibfnamefont {J.~W.}\ \bibnamefont
  {Olness}},\ and\ \bibinfo {author} {\bibfnamefont {E.~K.}\ \bibnamefont
  {Warburton}},\ }\bibfield  {title} {\bibinfo {title} {Lifetime of the
  $5^{-}$, 4585 $\mathrm{keV}$ $^{38}\mathrm{Ar}$ level},\ }\href
  {https://doi.org/10.1016/0370-2693(70)90253-4} {\bibfield  {journal}
  {\bibinfo  {journal} {Phys. Lett. B}\ }\textbf {\bibinfo {volume} {33B}},\
  \bibinfo {pages} {353} (\bibinfo {year} {1970})}\BibitemShut {NoStop}%
\bibitem [{\citenamefont {Kolata}(1976)}]{ref:Kolata1976}%
  \BibitemOpen
  \bibfield  {author} {\bibinfo {author} {\bibfnamefont {J.~J.}\ \bibnamefont
  {Kolata}},\ }\bibfield  {title} {\bibinfo {title} {Investigation of high-spin
  states in $^{38}\textrm{Ar}$},\ }\href
  {https://doi.org/10.1103/PhysRevC.13.1944} {\bibfield  {journal} {\bibinfo
  {journal} {Phys. Rev. C}\ }\textbf {\bibinfo {volume} {13}},\ \bibinfo
  {pages} {1944} (\bibinfo {year} {1976})}\BibitemShut {NoStop}%
\end{thebibliography}%

\end{document}